\documentclass[journal=ancac3,manuscript=article]{achemso}
\usepackage[version=3]{mhchem}
\usepackage{amssymb,epsfig,setspace,color}

\graphicspath{{figs/}}

\newcommand{\be}{\begin{equation}}
\newcommand{\ee}{\end{equation}}
\newcommand{\bg}{\begin{equation}}
\newcommand{\eg}{\end{equation}}
\newcommand{\bdm}{\begin{displaymath}}
\newcommand{\edm}{\end{displaymath}}
\newcommand{\bea}{\begin{eqnarray}}
\newcommand{\eea}{\end{eqnarray}}
\newcommand{\beas}{\begin{eqnarray*}}
\newcommand{\eeas}{\end{eqnarray*}}
\newcommand{\ba}{\begin{array}}
\newcommand{\ea}{\end{array}}
\newcommand{\nn}{\nonumber}

\newcommand{\bfg}{\begin{figure}}
\newcommand{\efg}{\end{figure}}

\newcommand{\fr}{\frac}

\newtheorem{lm}{Lemma}
\newtheorem{cl}{Corollary}
\newtheorem{df}{Definition}

\newcommand{\blm}{\begin{lm}}
\newcommand{\elm}{\end{lm}}
\newcommand{\bcl}{\begin{cl}}
\newcommand{\ecl}{\end{cl}}
\newcommand{\bdf}{\begin{df}}
\newcommand{\edf}{\end{df}}
\newcommand{\brk}{\begin{rm}}
\newcommand{\erk}{\end{rm}}

\newcommand{\lb}{\label}

\newcommand{\om}{\omega}

\newcommand{\Dt}{\Delta}

\newcommand{\veps}{\varepsilon}

\newcommand{\ld}{\lambda}

\newcommand{\gm}{\gamma}

\newcommand{\sg}{\sigma}

\newcommand{\ct}{\cite}
\newcommand{\rf}{\ref}

\newcommand{\vE}{{\bf E}}

\newcommand{\vr}{{\bf r}}

\title{Remarkable Predictive Power of the Modified Long Wavelength Approximation}

\author{Ilia L. Rasskazov}
\affiliation{The Institute of Optics, University of Rochester, Rochester, NY 14627, USA}

\author{Vadim I. Zakomirnyi}
\affiliation{Siberian Federal University, Krasnoyarsk, 660041, Russia}
\alsoaffiliation{Institute of Computational Modelling of the Siberian Branch of the Russian Academy of Sciences, Krasnoyarsk, 660036, Russia}
\email{vadimza@icm.krasn.ru}

\author{Anton D. Utyushev}
\affiliation{Siberian Federal University, Krasnoyarsk, 660041, Russia}
\alsoaffiliation{Institute of Computational Modelling of the Siberian Branch of the Russian Academy of Sciences, Krasnoyarsk, 660036, Russia}

\author{P. Scott Carney}
\affiliation{The Institute of Optics, University of Rochester, Rochester, NY 14627, USA}

\author{Alexander Moroz}
\affiliation{Wave-scattering.com}
\email{wavescattering@yahoo.com}

\begin{document}

\begin{abstract}
The modified long-wavelength approximation (MLWA), a next order approximation beyond the Rayleigh limit, has been applied usually only to the dipole $\ell=1$ contribution and for the 
range of size parameters $x$ not exceeding $x\lesssim 1$ to estimate far- and near-field electromagnetic properties of plasmonic nanoparticles. Provided that the MLWA functional form for the $T$-matrix in a given channel $\ell$ is limited to the ratio $T\sim iR/(F+D-iR)$, where 
$F$ is the familiar size-independent Fr\"ohlich term and $R\sim {\cal O}(x^{2\ell+1})$ is a radiative reaction term, there is a one-parameter freedom in selecting the dynamic 
depolarization term $D\sim {\cal O}(x^2)$ which preserves the fundamental feature of the 
MLWA that its predictions coincide with those of the Mie theory up to the order ${\cal O}(x^2)$.
By exploiting this untapped design freedom, we demonstrate on a number of different metals 
(Ag, Al, Au, Mg), and using real material data, that the MLWA may surprisingly yield very 
accurate results for plasmonic spheres both for (i) $x$ up to $\gtrsim 1$ and beyond, 
and (ii) higher order multipoles ($\ell>1$), essentially doubling its expected range 
of validity. Because the MLWA obviates 
the need of using spherical Bessel and Hankel functions and allows for 
an intuitive description of (nano)particle properties
in terms of a driven damped harmonic oscillator parameters, a
significantly improved analysis and understanding of nanoparticle scattering and near-field 
properties can be achieved.
\end{abstract}

\section{Introduction}

For over a century, the Mie theory~\cite{Mie1908}, which provides a rigorous and complete 
description of the light scattering from spherical particles, has been an indispensable tool for 
numerous applications.
The theory offers a reliable description of electromagnetic fields inside, in a close proximity of, 
or far away from spherical particles of any material, and has been employed in ongoing development 
of plasmonic applications of small metal particles in biology, energy conversion, medicine, sensing, 
and many other fields. 
Nonetheless, the popularity of intuitive approximate descriptions of light scattering 
nowadays appears to be even greater than before, as witnessed by applications of coupled mode 
theory~\cite{Hamam2007,Ruan2010a,Ruan2011}, coupled harmonic oscillators model \cite{Joe2006,Gallinet2011,Limonov2017}, 
and modal expansions \cite{Grigoriev2013a,Colom2017,Colom2019a}. 
The prime example is the {\em Rayleigh limit}, 
which preceded the Mie theory~\cite{Mie1908} and, in spite of its 
deficiencies (e.g. cannot account for a size dependent red-shift of the localized 
surface plasmon resonance (LSPR) 
(cf. sec 12.1.1. of ref \citenum{Bohren1998}) and predicts zero extinction and negative absorption 
for purely dielectric particles), 
it has been repeatedly used in various analytic considerations to describe LSPR, 
the basics of surface-enhanced Raman spectroscopy (SERS)~\cite{Moskovits1985} and other 
plasmonic properties. 

The focus of the present work is on the so-called modified long-wavelength approximation 
(MLWA) \cite{Meier1983,Zeman1984,Zeman1987,Kelly2003,Kuwata2003,Moroz2009,Zoric2011,Massa2013,LeRu2013,Schebarchov2013,
Januar2020,Rasskazov20OL}, 
which is a next-order approximation beyond the Rayleigh limit. The present study is limited to spherical particles, in which case the MLWA yields results coinciding with 
the Mie theory up to the order ${\cal O}(x^2)$, with $x=k r_s$ being the conventional 
size parameter, where $r_s$ is spherical particle radius and $k$ is 
wave number of an incident plane wave in the host medium.
The MLWA overcomes the main shortcoming of the quasi-static Rayleigh limit and has been known to rather 
precisely capture {\em size-dependence} of the elementary cross sections, 
including the red-shift of the LSPR (sec 12.1.1. of ref \citenum{Bohren1998}), 
an additional red-shift of the near-field intensity peak relative to the LSPR 
\cite{Januar2020}, and local field enhancements. 
Usually, like the Rayleigh limit, the MLWA was, with a notable exception~\cite{Schebarchov2013}, 
applied only to the electric dipole ($\ell=1$) 
contribution. The results of refs~\citenum{Schebarchov2013,Rasskazov20OL} have shown 
that the MLWA can be surprisingly accurate not only for $x\lesssim 1$ 
(which is the expected range of validity of the MLWA by its very definition), but 
also for higher order multipole contributions ($\ell>1$), and for $x$ well above 
unity~\ct{Schebarchov2013,Rasskazov20OL}. The unexpected accuracy of the MLWA 
is a pleasant and very useful surprise that deserves further examination. 

Schebarchov et al.~\ct{Schebarchov2013} needed to keep terms up to the order 
${\cal O}(x^4)$ in the expansion of spherical Bessel functions to achieve
reliable results for $x\gtrsim 1$.
Surprisingly enough, we show here that one can improve 
on the predictive power of the ${\cal O}(x^4)$ approximation 
of ref~\citenum{Schebarchov2013} by keeping only terms ${\cal O}(x^2)$ 
upon optimizing the so-called {\em dynamic depolarization} term.
Our focus on the ${\cal O}(x^2)$-MLWA is that, at least in the dipole case, 
only the ${\cal O}(x^2)$-MLWA allows for an intuitive
description of scattering and near-field properties of
Drude-like plasmonic particles in terms of a driven damped harmonic oscillator 
with the Abraham-Lorenz force, mass, and stiffness directly related to
corresponding partial depolarization terms \cite{Januar2020}.
Such an intuitive description is easy to analyze, which
may result in a significantly improved understanding of nanoparticle scattering 
and its near-field, thereby facilitating design of (nano)particles with desired properties.
Our optimized ${\cal O}(x^2)$-MLWA captures photonic properties
of a particle as an interplay of three basic terms: a quasi-static {\em Fr\"ohlich} term $F$, 
the dynamic depolarization term $D$ ($\sim x^2$), and a {\em radiative reaction} 
term $R$ ($\sim x^{2\ell+1}$), while providing a better match to the Mie theory than that obtained
in earlier works \ct{Schebarchov2013,Rasskazov20OL}. We demonstrate this in detail 
for various plasmonic particles (Ag, Al, Au, Mg) for $\ell\ge 1$ and beyond $x=1$ on using 
real material data \cite{McPeak2015,Palm2018}.
The optimized MLWA will be shown to yield remarkable agreement of the 
peak position and height with those 
in the exact Mie theory, and to possess 
an enlarged range of validity essentially twice as large as it has been initially expected. 

Particle electric multipole polarizabilities $\alpha_\ell$ can be obtained 
in the quasi-static limit from the corresponding $T$-matrix elements $T_{E\ell}$ conveniently described by MLWA. Therefore, 
any quasi-static method intrinsically based on elementary polarizabilities, 
such as Maxwell-Garnett homogenization formulas \cite{Yannopapas2005,Markel2016,Markel2016a}, a coupled-dipole (CDA) or 
discrete-dipole approximation (DDA)~\cite{Purcell1973,Draine1994,Yurkin2007}, 
Gersten and Nitzan approximation for determining nonradiative decay rates 
\cite{Gersten1981,Moroz2010} can, in principle, be immediately improved by adopting our results.
An insight provided by our MLWA can be also straightforwardly employed in layer and bulk
Korringa-Kohn-Rostocker photonic multiple-scattering theories 
\cite{Korringa1947,Kohn1954,Moroz1995,Moroz1999,Stefanou1998}
to analyze in simple terms the effect of periodic arrangement 
of spherical scatterers in a plane, or in a three-dimensional lattice, 
on various single-sphere multipole contributions.

\section{Notation, Definitions, and the Rayleigh Limit}
\lb{sc:def}
The resulting cross sections for a plane electromagnetic wave scattering 
from a spherical particle of radius $r_s$ are given as an infinite sum over 
all momentum channels $\ell\ge 1$ and both polarizations~\cite{Newton1982,Bohren1998}.
According to eqs 2.135-8 of ref~\citenum{Newton1982}, any given angular momentum 
channel $\ell$ and polarization $p$ ($p=E$ for electric (or TM) 
polarization, and $p=M$ for magnetic (or TE) polarization) contributes the 
following partial amount to the resulting scattering, absorption, and extinction cross sections,

\begin{eqnarray}
\sigma_{sca;p\ell} &= & \frac{2(2\ell +1)\pi}{k^2}\, |T_{p\ell}|^2,
\label{sgsc}
\\
\sigma_{abs;p\ell} &= & - \frac{2(2\ell +1)\pi}{k^2}\, \left[|T_{p\ell}|^2 +
\Re(T_{p\ell}) \right],
\label{sgabs}
\\
\sigma_{ext;p\ell} &= & -\frac{2(2\ell +1)\pi}{k^2}\, \Re(T_{p\ell}),
\label{sgtot}
\end{eqnarray}
where $k=2\pi/\ld$ is the wavenumber, with $\ld$ being the incident 
wavelength in the \textit{host medium}. In an optical convention, $-T_{p\ell}$ are nothing 
but familiar Mie's expansion coefficients 
$a_\ell$ and $b_\ell$ (eqs 4.53 of ref~\citenum{Bohren1998}), i.e. $T_{E\ell} = -a_\ell$ 
and $T_{M\ell} = -b_\ell$. 
The resulting full cross sections are determined as an infinite sum
\begin{equation*}
 \sigma_{sca} = \sum_{p,\ell} \sigma_{sca;p\ell} \ , \quad 
 \sigma_{abs} = \sum_{p,\ell} \sigma_{abs;p\ell} \ , \quad
 \sigma_{ext} = \sum_{p,\ell} \sigma_{ext;p\ell}.
\lb{csisum}
\end{equation*}

In the case of a homogeneous sphere, Mie solution \cite{Mie1908} 
explicitly determines the respective $T$-matrix elements in a given 
$\ell$th angular momentum channel as (eqs 2.127 of ref~\citenum{Newton1982}) 
\begin{equation}
T_{p\ell} = - \frac{ \upsilon [xj_\ell(x)]' j_\ell(x_s) 
 - j_\ell(x) [x_sj_\ell(x_s)]'}
 { \upsilon [x h_\ell(x)]' j_\ell(x_s) 
 - h_\ell(x) [x_sj_\ell(x_s)]'}, 
\label{miecoef}
\end{equation}
where the respective $j_\ell$ and $h_\ell=h_\ell^{(1)}$ are the conventional spherical Bessel and Hankel 
functions 
(sec 10 of ref \citenum{Abramowitz1973}), prime denotes the derivative with 
respect to the argument,
$x=k r_s$ is the conventional dimensionless size parameter, $x_s=x n$, where $n=n_s/n_h=\sqrt{\veps_s\mu_s/(\veps_h\mu_h)}$ 
is the relative refractive index contrast of the sphere ($n_s$) and the host ($n_h$), and 
$\veps_s$ ($\veps_h$) is the sphere (host) permittivity, and $\mu_s$ ($\mu_h$) is the sphere (host) permeability.
One has, assuming nonmagnetic media, $\upsilon=\mu_s/\mu_h=1$ for magnetic polarization, and $\upsilon=\veps:=\veps_s/\veps_h$ 
for electric polarization. 

In the familiar Rayleigh limit,
\begin{equation}
T_{E1} \to T_{E1;R}= \fr{2ix^3}{3} \frac{\veps-1}{\veps+2} \qquad (x\ll 1).
\label{sphrp}
\end{equation}
The overall $x^3$-factor of $T_{E1;R}$ in eq \rf{sphrp} affects only the magnitude of 
$T_{E1;R}$, but not its structure (i.e. a peak position). 
Because $T_{E1;R}$ is {\em purely imaginary} for 
real $\veps_s$, the resulting extinction cross section $\sigma_{ext;E1}$ 
for a purely dielectric sphere is, because $\Re(T_{E1;R})=0$ in eq \rf{sgtot}, 
identically zero in the Rayleigh limit. Therefore the rigorous bound $-\Re(T_{E1;R})\ge |T_{E1;R}|^2$
is not satisfied, which violates unitarity (eqs 5a-b of ref~\citenum{Chylek1979} and eq C6 of ref~\citenum{Moroz2009}). 
Moreover, the resulting absorption cross section 
$\sigma_{abs;E1}$ for purely dielectric, and hence nonabsorbing, sphere is, 
in virtue of $\Re(T_{E1;R})=0$ in eq \rf{sgabs}, {\em negative}, $\sigma_{abs;E1}=-\sigma_{sca;E1}<0$.
The modulus of $T_{E1;R}$ is also not prevented from exceeding unity, thus violating rigorous bound $|T_{p\ell}|\le 1$ and
implying complex phase shifts with wrong imaginary part for $\veps\approx -2$.
From the above fundamental perspective, the Rayleigh limit fails and it is a wonder that 
the Rayleigh limit is used so often and in so many different settings~\cite{Moskovits1985,Fan2014,Yezekyan2020}.

\section{Taming the Zoo of MLWA's}
\lb{sc:zoo}
The usual MLWA \ct{Meier1983,Kelly2003,Moroz2009,LeRu2013,Schebarchov2013} is a limiting form of the Mie 
dipole term for $x\lesssim 1$, that, unlike the quasi-static Rayleigh approximation (eq \rf{sphrp}), 
keeps both dynamic depolarization ($D\sim x^2$) and radiative reaction ($R\sim x^3$) terms.
The MLWA for a general $\ell$ and $p$ is 
required to reproduce the Mie theory results up to the order ${\cal O}(x^2)$. It 
combines in a concise way three different elementary terms,
involving the dynamic depolarization ($D\sim x^2$) and 
radiative reaction ($R\sim x^{2\ell +1}$ for $p=E$ and 
$R\sim x^{2\ell +3}$ for $p=M$), in the functional form
\bg
T_{p\ell} \sim \fr{i R(x)} {F + D(x) -i R(x)},
 \lb{mlwaff}
\eg
where
\bg
F := \upsilon + \fr{\ell+1}{\ell} 
\lb{FT}
\eg
is a {\em size-independent} quasi-static Fr\"ohlich term.
Without going yet in to the details of $D$ and $R$, the sole functional form (eq \ref{mlwaff})
makes it already transparent that the usual Rayleigh limit (eq \rf{sphrp}), which 
amounts to setting $D(x)=R(x)\equiv 0$ in the denominator, 
is essentially recovered for $x, x_s\ll 1$, $\ell=1$, and $p=E$.
The vanishing of the size-independent $F$ in the denominator yields the usual 
quasi-static Fr\"ohlich LSPR condition, which determines the 
quasi-static LSPR frequencies $\om_{0\ell}$. In the case of Drude fit of $\veps_s$,
\begin{equation*}
 \veps_s = \veps_\infty - \dfrac{\om_p^2}{\om (\om + i \gamma) } ,
\lb{dfit}
\end{equation*}
where $\veps_\infty$ is the high-frequency permittivity limit, 
$\om_p$ is the bulk plasma frequency, and $\gm$ is a damping constant, 
one finds $\om_{0\ell}=\om_p/\sqrt{\veps_\infty+[(\ell+1)\veps_h/\ell]}$. 
A size-dependent red shift of the dipole LSPR, which cannot be accounted for by the Rayleigh approximation,
is determined by solving for the zeros of the sum $F + D(x)=0$.
For purely {\em real} $\upsilon$ all the terms $F, D, R$ are real.
Although the order of $R$ of at least $\sim x^{2\ell+1}$ in the denominator is larger than that of $x^2$, 
it is its presence there which ensures for purely real $\upsilon$ that 
$|T_{p\ell}|^2= -\Re(T_{p\ell})$, i.e. $\sigma_{abs;\ell}\equiv 0$ 
(cf. eq \ref{sgabs}), and unitarity 
$\sigma_{ext;\ell}= \sigma_{sca;\ell}$ (cf. eqs \ref{sgsc}, \ref{sgtot}).
Furthermore, keeping $R(x)$ in the denominator prevents the modulus of $T_{p\ell}$ from exceeding unity 
for $x\gtrsim 1$, complying with the rigorous bound $|T_{p\ell}|\le 1$.

The restriction $\sigma_{abs;\ell}\ge 0$ translates on substituting
eq \ref{mlwaff} into eq \ref{sgabs} in the constraint
\bg
\Im \left\{ R(x) [F^* + D^*(x)]\right\}\ge 0.
\label{sgabsc}
\eg

Rigorously speaking, provided that we perform the limit $x, x_s\ll 1$ in each of the numerator and denominator
of the $T$-matrix (eq \rf{miecoef}) following the recipe that
\begin{itemize}

\item[({\bf R1})] only terms up to ${\cal O}(x^2)$ order in the asymptotic expansion of each spherical 
Bessel function in eq \rf{miecoef} are kept (see chapter 10 of ref \citenum{Olver2010})

\item[({\bf R2})] in a product of a spherical Bessel and Hankel 
functions, or of two Bessel functions, again only terms up to 
${\cal O}(x^2)$ order are kept and all higher order terms are neglected

\end{itemize}
any such a limit functional form (eq \ref{mlwaff}) is {\em ambiguous} (see the Supporting Information).
For example, one can arrive in a given $(E\ell)$ channel for 
nonmagnetic media characterized by $\veps=n^2$ at (see the Supporting Information)
\bea
T_{E\ell} &\sim& \fr{i R_{E\ell}(x) \left[1 + \fr{\veps x^2}{(\ell+1)(2\ell+3)}\right]}
 {
F + \fr{\veps}{(\ell+1)(2\ell+3)} 
 \left[\veps -\fr{(\ell+1)(2\ell+3)}{\ell(2\ell-1)} \right]
 \, x^2 - i R_{E\ell}(x)}
\lb{mlwa1}
\\
 &\sim& \fr{i R_{E\ell}(x)}
 {
 F - \fr{2(2\ell+1)}{\ell(2\ell-1)(2\ell+3)} \, \veps x^2 
-i R_{E\ell}(x)}
\lb{mlwa2}
\\
 &\sim&
\fr{i R_{E\ell}(x) \left[1-(\veps+1)\,\fr{x^2}{2(2\ell+3)}\right]}
{
F +\left[-\veps^2 -\fr{3(2\ell+1)}{\ell(2\ell-1)}\,\veps
+\fr{(\ell+1)(2\ell+3)}{\ell(2\ell-1)}\right]\fr{x^2}{2(2\ell+3)}
- i R_{E\ell}(x)}
\lb{mlwa3}
\\
 &\sim&
\fr{i R_{E\ell}(x)}
{
F +\left(\fr{\ell-2}{\ell+1}\, \veps + 1\right) 
\fr{(\ell+1)(2\ell+1)}{\ell(2\ell-1)(2\ell+3)}\, x^2
- i R_{E\ell}(x)},
\lb{tmtflr3c}
\eea
where 
\bg
R_{E\ell}(x) :=
\fr{(\veps-1)\, (\ell+1) x^{2\ell+1}}{\ell (2\ell-1)!! (2\ell+1)!!}\cdot
\lb{radcr}
\eg
All the above eqs \rf{mlwa1}--\rf{tmtflr3c} constitute legitimate MLWA's. 
The expression \rf{mlwa1} results from 
(i) first factorizing the $T$-matrix in terms of the so-called $K$-matrix 
and (ii) taking the limit $x, x_s\ll 1$ in the $K$-matrix (see the Supporting Information). 
The expression \rf{mlwa3} results from 
taking the limit $x, x_s\ll 1$ directly in the numerator and denominator of the $T$-matrix in eq \rf{miecoef}. Provided that
the numerator of the $T$-matrix has the form $R_{E\ell}[1+Ax^2]$
as in eqs \rf{mlwa1}, \rf{mlwa3}, following our recipe ({\bf R1})-({\bf R2}) we can either ignore the ${\cal O}(x^2)$ term
in the square bracket in therein, or we may decide to
multiply the numerator and denominator of any of 
the above expression (eqs \rf{mlwa1}, \rf{mlwa3}) by $f=1 -A x^2$. 
The latter provision will transform the ${\cal O}(x^2)$ term in the numerator 
into ${\cal O}(x^4)$ term, which is subsequently ignored. However, this provision
brings about a change of $D$ in the denominator.
This way one arrives at further alternative MLWA expressions. 
For example, eq \rf{mlwa2} is obtained from eq \rf{mlwa1} by multiplying both the numerator 
and denominator of the rhs of eq \rf{mlwa1} by $1-\veps x^2/[(\ell+1)(2\ell+3)]$. $D(x)$ in eq \rf{tmtflr3c} is obtained from eq \rf{mlwa3} by multiplying 
both the numerator and denominator of the latter by $1+(\veps+1) x^2/[2(2\ell+3)]$. 
\begin{figure}
\centering
\includegraphics[width=3.33in]{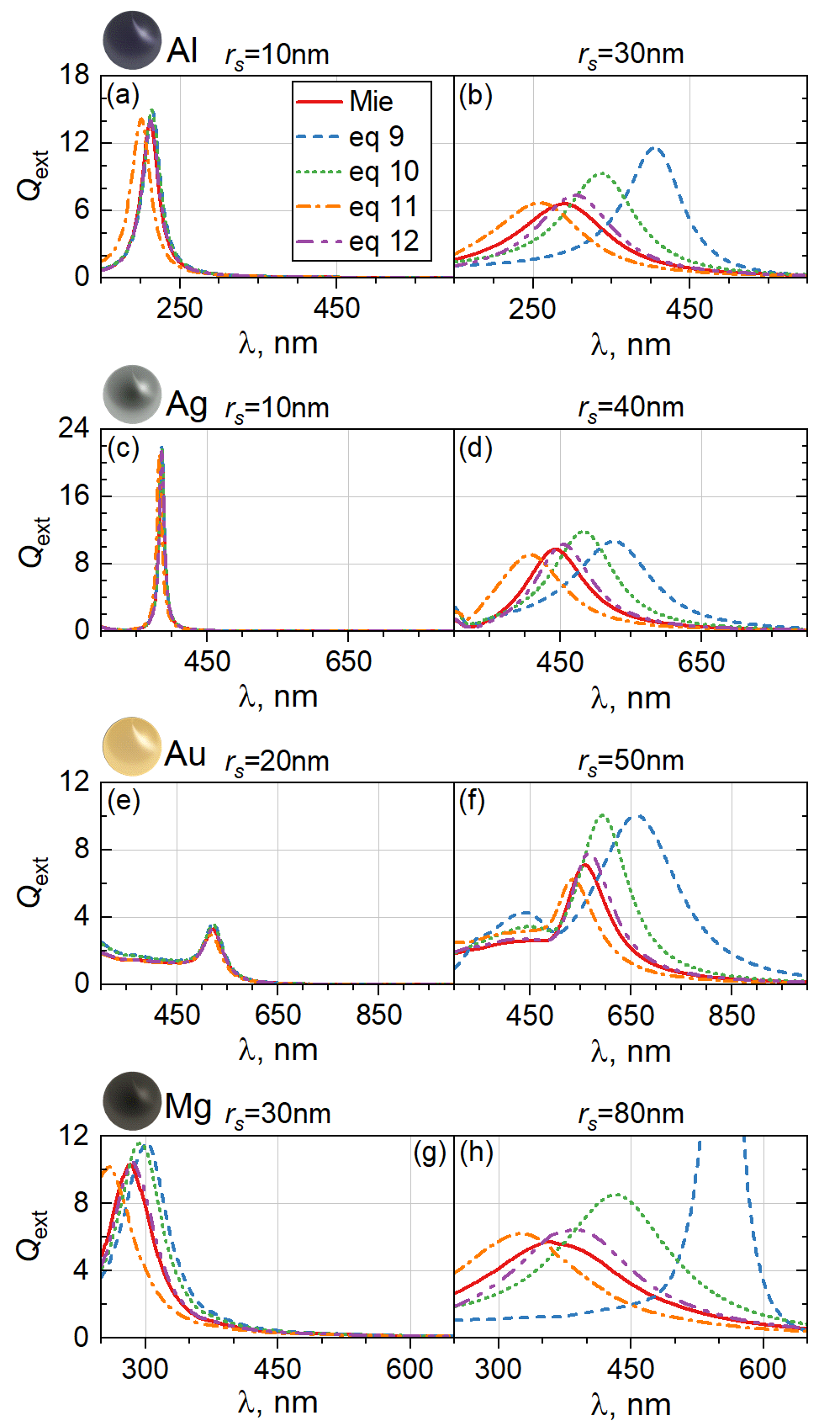}
\caption{Comparison of extinction spectra for 
(a), (b) Al,
(c), (d) Ag,
(e), (f) Au,
(g), (h) Mg NPs with different radii $r_s$ in water ($n_h=1.33$) host
as calculated for electric dipole, $(E1)$ channel, via exact Mie theory, eq~\ref{miecoef}, 
and via different limiting expressions for 
MLWA: eqs~\ref{mlwa1}--\ref{tmtflr3c}.
Notice almost the same accuracy of MLWA's for small $r_s$ (left), and different accuracy for large $r_s$ (right).
Real material data is used for Al, Ag, Au~\cite{McPeak2015} and Mg~\cite{Palm2018}. 
}
\label{fig:zoo}
\end{figure}
All expressions \rf{mlwa1}--\rf{tmtflr3c} 
make it transparent that $T_{E\ell}$ in any given channel is determined solely by a size-independent quasi-static Fr\"ohlich term $F$ (eq \rf{FT}), a {\em dynamic depolarization} term 
$D$ ($\sim x^2$), and a {\em radiative reaction} term $R_{E\ell}$ (eq \rf{radcr}). 
However, as illustrated in Figure~\ref{fig:zoo}, MLWA's predictions may dramatically differ 
for $x, x_s\gtrsim 1$, in spite of that they all agree for $x, x_s\ll 1$. As obvious from Figure \rf{fig:zoo}, the MLWA
of eq \rf{tmtflr3c} stands out by its performance from $x\ll 1$ up to $x\gtrsim 1$~\cite{Schebarchov2013,Rasskazov20OL}.
Nevertheless, all MLWA's correctly account up to ${\cal O} (x^2)$-term for a size-dependent red shift of the dipole LSPR, 
whereas the Rayleigh approximation does not. 
For instance, for $\ell=1$, eq \ref{tmtflr3c} becomes (cf. eq A3 of ref \citenum{Moroz2009})
\bea
T_{E1} &\sim& \fr{2ix^3}{3} \fr{(\veps-1)} 
{\veps + 2-\fr{3}{5}(\veps -2)\, x^2 -\fr{2i}{3}(\veps-1)\, x^3}\cdot
 \lb{tmtflr3c1}
\eea
On substituting into eqs \ref{sgsc}--\ref{sgtot}, one finds the following cross sections of the dipole MLWA contribution:
\bea
\sg_{sca;E1} &=& \fr{4\pi}{15 k^2}\, 
\frac{10\, x^6 \left| \veps-1\right|^2}{\left|\veps+2
-\fr{3}{5} (\veps-2) x^2-i \fr{2}{3} (\veps-1) x^3\right|^2},
\lb{sgscsdm}
\\
\sg_{abs;E1} &=& \fr{4\pi}{15 k^2}\, \frac{9\, x^3 \left(x^2+5\right) \Im(\veps)}{\left|
\veps+2-\fr{3}{5} (\veps-2) x^2-i \fr{2}{3} (\veps-1) x^3\right|^2},
\lb{sgabsdm}
\\
\sg_{ext;E1} &=& \fr{4\pi}{15 k^2}\, \frac{9\, x^3 \left(x^2+5\right) 
\Im(\veps)+10\, x^6 \left|\veps-1\right|^2}
{\left|\veps+2-\fr{3}{5} (\veps-2) x^2-i \fr{2}{3} (\veps-1) x^3\right|^2}\cdot
\lb{sgtotsdm}
\eea
The unitarity of the MLWA, i.e. that $\sigma_{ext;p\ell}=\sigma_{sca;p\ell}+\sigma_{abs;p\ell}$,
can be easily checked. One can also easily verify that, for $\Im(\veps)=0$, 
the common denominator $|\Dt|^2$, 
$\Dt(x):= F + D(x) -i R(x)$, of the dipole MLWA cross sections (eqs \rf{sgscsdm}--\rf{sgtotsdm}) vanishes at
\bg
\Re(\veps) \approx -2-\fr{12 x^2}{5} 
\lb{vepsz}
\eg
up to the order $x^2$, in which case $\Dt\approx {\cal O} (x^3)$. 
The latter is, as it should, in agreement with the dipolar LSPR position in the exact Mie theory 
up to the order of $x^2$ (see sec 12.1.1 of ref~\citenum{Bohren1998}).
The very same red-shift (eq \rf{vepsz}) follows, as it should, also on using any of eqs \ref{mlwa1}--\ref{tmtflr3c}. 
The easiest way to verify this is to substitute eq~\rf{vepsz} into the respective denominators of eqs \ref{mlwa1}--\ref{tmtflr3c}, whereby the real part of each
denominator vanishes including the terms ${\cal O} (x^2)$.
For a general $(E\ell)$-pole, the position of an $(E\ell)$-pole LSPR up to the order $x^2$ can be implicitly found as
\bg
\Re(\veps) \approx -\fr{\ell+1}{\ell} 
-\fr{2(\ell+1)(2\ell+1)}{\ell(2\ell-1)(2\ell+3)}\, x^2.
\lb{vepszl}
\eg

In order to better address the above intrinsic ambiguity of the MLWA, let us limit its definition as an approximation which satisfies the following axioms:
\begin{itemize}

\item[({\bf A1})] It has the functional form (eq \ref{mlwaff}), with the Fr\"ohlich term $F$ and the radiative reaction term $R$ fixed by eqs \rf{FT} and \rf{radcr}, respectively. 

\item[({\bf A2})] A dynamic depolarization term 
$D$ ($\sim x^2$), required to be at most linear in $\veps$, has to reproduce the red shift (eq \rf{vepszl}).

\end{itemize}
The above axioms are obviously satisfied by the MLWA's (eq \rf{mlwa2}) and (eq \ref{tmtflr3c}). 
It is not difficult to demonstrate that for each $\ell$ there is an infinite one parameter 
continuous family of MLWA's, which satisfy the above axioms {\bf A1}-{\bf A2}, yet they have all different dynamic depolarization term. Just assume $D$ in the form
\bg
D=(\mathfrak{a}\veps +\mathfrak{b}) x^2, \qquad \mathfrak{b}=\fr{\ell+1}{\ell} \, \mathfrak{a}+ \fr{2(\ell+1)(2\ell+1)}{\ell(2\ell-1)(2\ell+3)},
\lb{gDdf}
\eg
where $\mathfrak{a}, \mathfrak{b}$ are real numbers. Note that
\bea
\lefteqn{
\Im\left\{ (\veps-1) [\veps^* + \fr{\ell+1}{\ell} + (\mathfrak{a}\veps^* +\mathfrak{b})x^2 ]\right\}
}
\nn\\
&&=\Im\left\{\veps \fr{\ell+1}{\ell}- \veps^* + (-\mathfrak{a}\veps^* +\mathfrak{b}\veps)x^2 \right\}
= \left[\fr{(2\ell+1)}{\ell}+ (\mathfrak{a}+\mathfrak{b})x^2 \right] \Im (\veps),
\label{sgabscl}
\eea
where we have have used $\Im (-\veps^*)=\Im (\veps)$ to arrive at the last equality. In the case of passive media without any gain $\Im (\veps)\ge 0$. Thus the constraint \ref{sgabsc} is automatically satisfied for $\mathfrak{a}, \mathfrak{b}\ge 0$. In particular, when eq \ref{sgabscl} is combined with eq \ref{gDdf}, eq \ref{sgabscl} imposes a limitation on $x$ only for $\mathfrak{a}<0$, in which case the constraint becomes
\begin{equation*}
1 + \left[ \mathfrak{a}+\fr{2(\ell+1)}{(2\ell-1)(2\ell+3)} \right] x^2 \ge 0.
\label{sgabscll}
\end{equation*}
In the limit $\ell\gg 1$ this reduces to $1 + 
\mathfrak{a}x^2 \gtrsim 0$, which is in agreement with that
$\mathfrak{b}\approx \mathfrak{a}$ for $\ell\gg 1$.

As it has been illustrated in Figure~\ref{fig:zoo},
altering the dynamic depolarization term $D$, while of course obeying the axiom {\bf A2}, can have a very dramatic effect on the MLWA's predictions.
In what follows, we exploit the untapped one-parameter freedom in selecting an optimized $D$.

\section{MLWA with Optimized Dynamic Depolarization Term}
\begin{figure}[t]
 \centering
 \includegraphics{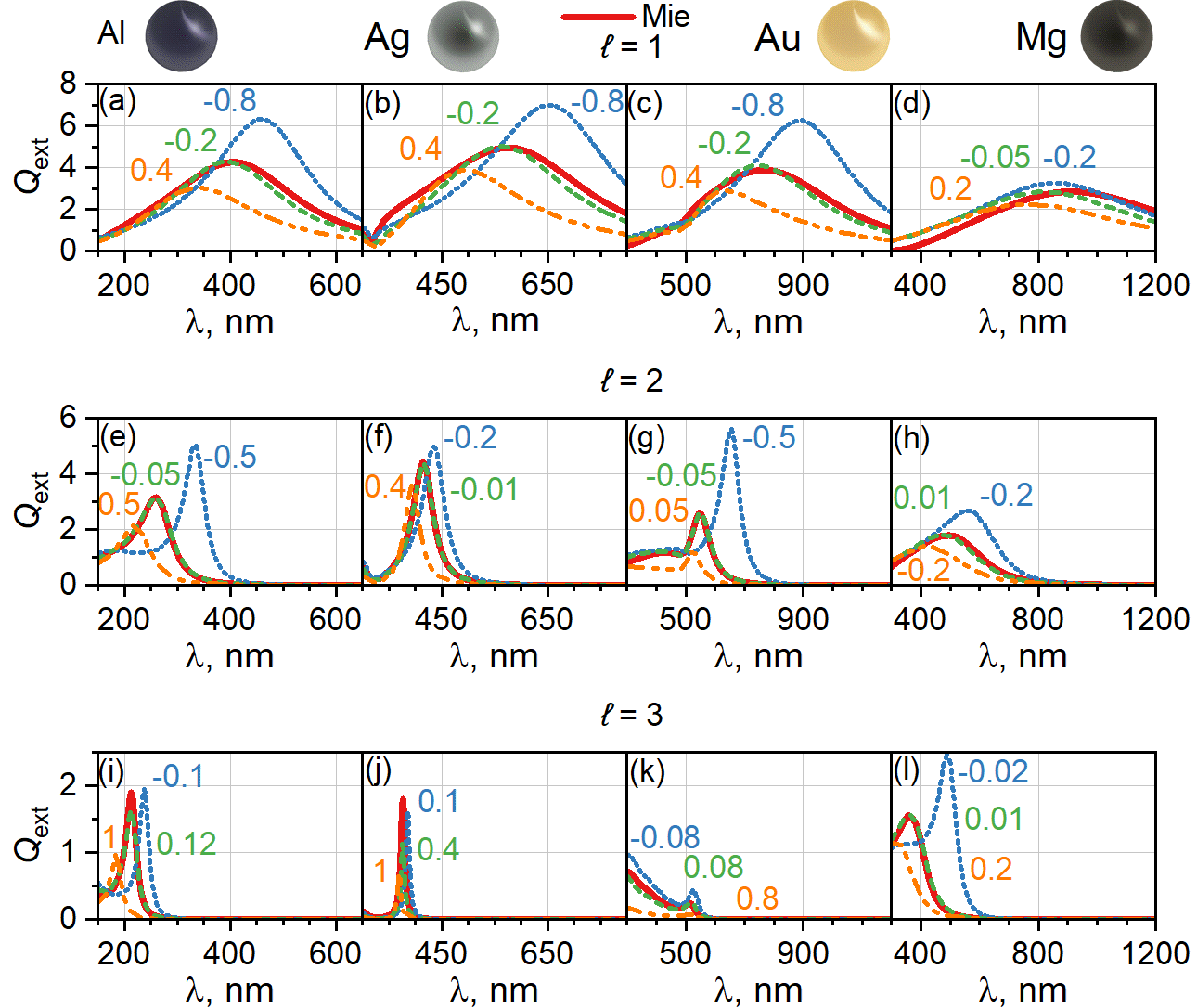}
 \caption{The effect of optimizing $D$ on the contributions of $\ell=1,2,3$ electric multipoles 
to the extinction efficiency of Al, Ag, Au and Mg nanospheres with $r_s=50$~nm, $r_s=70$~nm, $r_s=100$~nm, and $r_s=120$~nm, respectively.
 Spectra are calculated via exact Mie theory (eq~\ref{miecoef}) and via MLWA (eq \ref{mlwaff}) with dynamic depolarization from eq~\ref{gDdf} on using different $\mathfrak{a}$ as labeled in plots.}
 \label{fig:mlwa_a}
\end{figure}

\lb{sc:opt}
The results shown in Figure~\ref{fig:mlwa_a}
demonstrate the effect of optimizing $D$ on the contributions of $\ell=1,2,3$ electric multipoles 
to the extinction efficiency of Ag, Al, Au and Mg nanospheres using real material data.
The noble metals Ag and Au are traditional plasmonic materials, yet they are rare and expensive. 
Al is much cheaper plasmonic metal, with resonances in the UV and visible 
up to $\sim 700$~nm owing to a strong interband transition leading to high losses at lower energies~\cite{Langhammer2008,Ross2014a}. Mg is one of the recently emerged plasmonic metals having a broad operating range. Recent experimental
work has demonstrated that top-down fabricated Mg nanostructures sustain LSPRs \cite{Biggins2018,Ringe2020}. 

\begin{figure}
\centering
\includegraphics[width=3.33in]{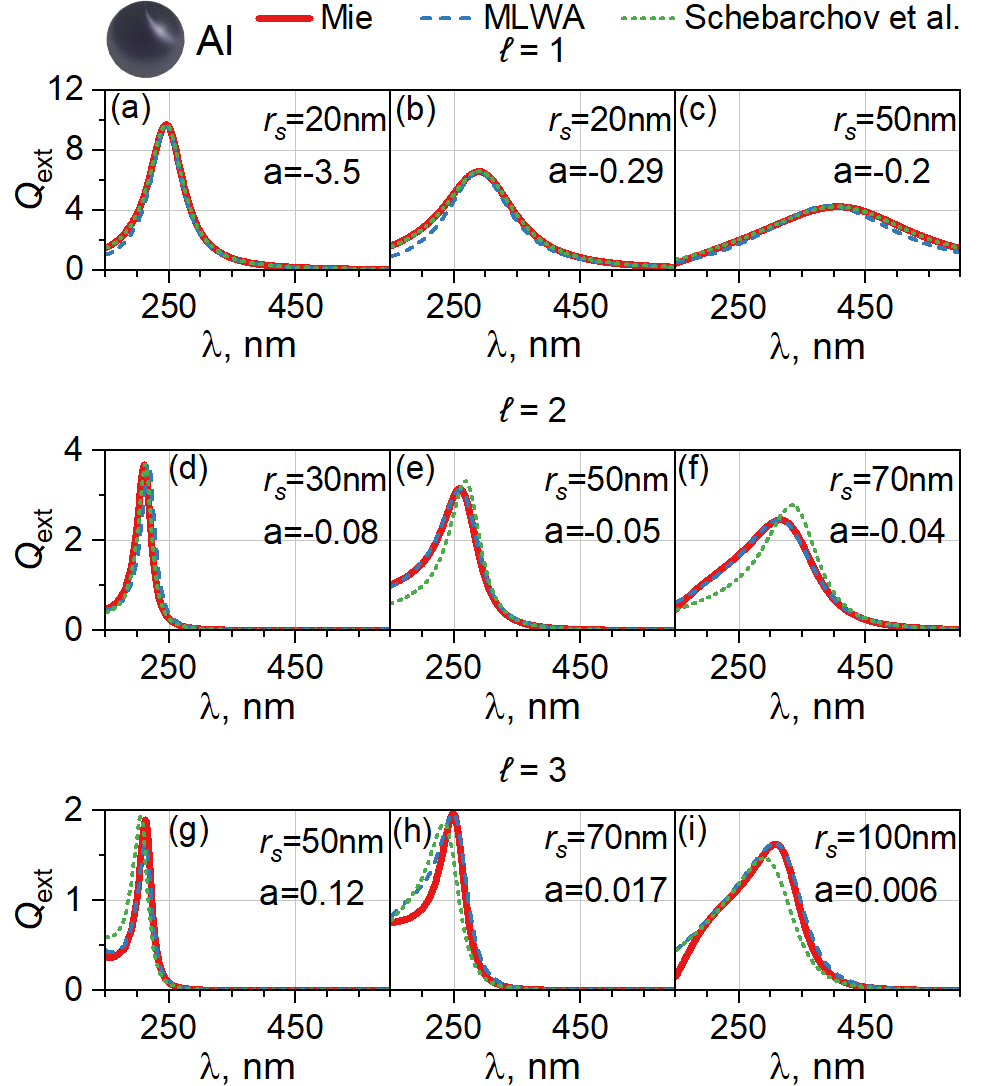}
\caption{Comparison of the contributions of $\ell=1,2,3$ electric multipoles to the extinction efficiency of 
Al nanospheres in water ($n_h=1.33$) host as calculated via exact Mie theory (eq \ref{miecoef}), 
via MLWA (eq \ref{mlwaff}) with dynamic depolarization from eq~\ref{gDdf} on using optimized parameter $\mathfrak{a}$ as labeled in plots, and 
via eqs 32, 39 and 40 of ref~\citenum{Schebarchov2013} for $\ell=1,2,3$, respectively.
Notice different $r_s$ are chosen for the most representative results.
}
\label{fig:Al_occam}
\end{figure}

\begin{figure}
\centering
\includegraphics[width=3.33in]{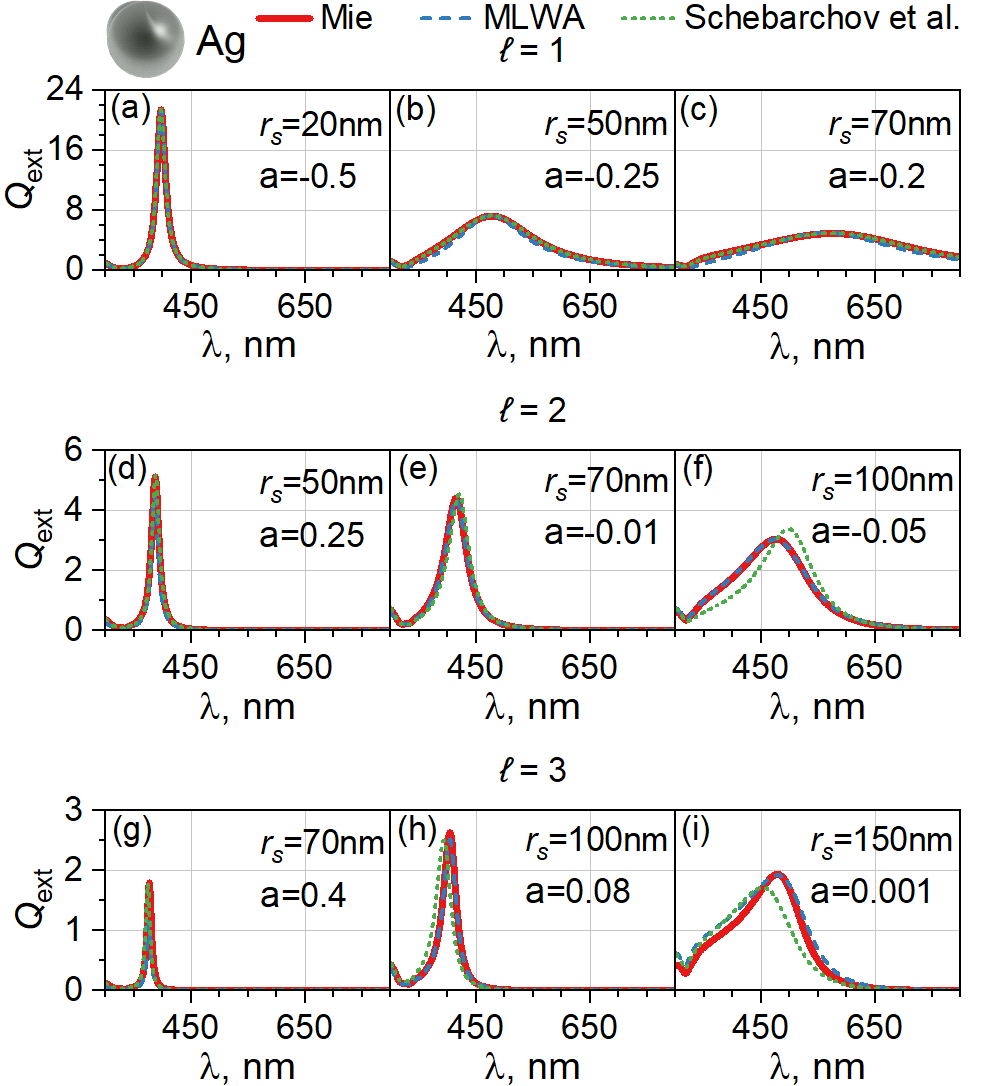}
\caption{The same as in Figure~\ref{fig:Al_occam}, but for Ag spheres.
}
\label{fig:Ag_occam}
\end{figure}

\begin{figure}
\centering
\includegraphics[width=3.33in]{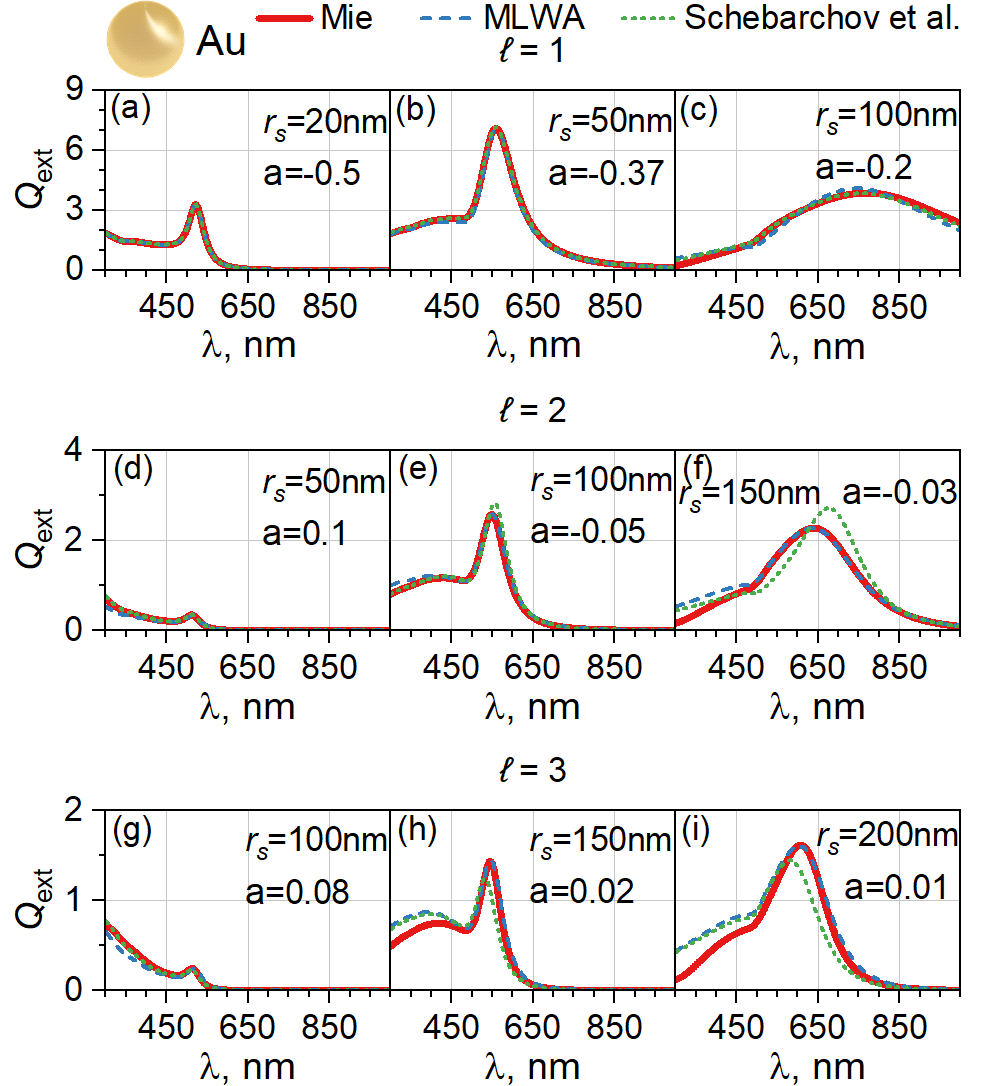}
\caption{The same as in Figure~\ref{fig:Al_occam}, but for Au spheres.
}
\label{fig:Au_occam}
\end{figure}

\begin{figure}
\centering
\includegraphics[width=3.33in]{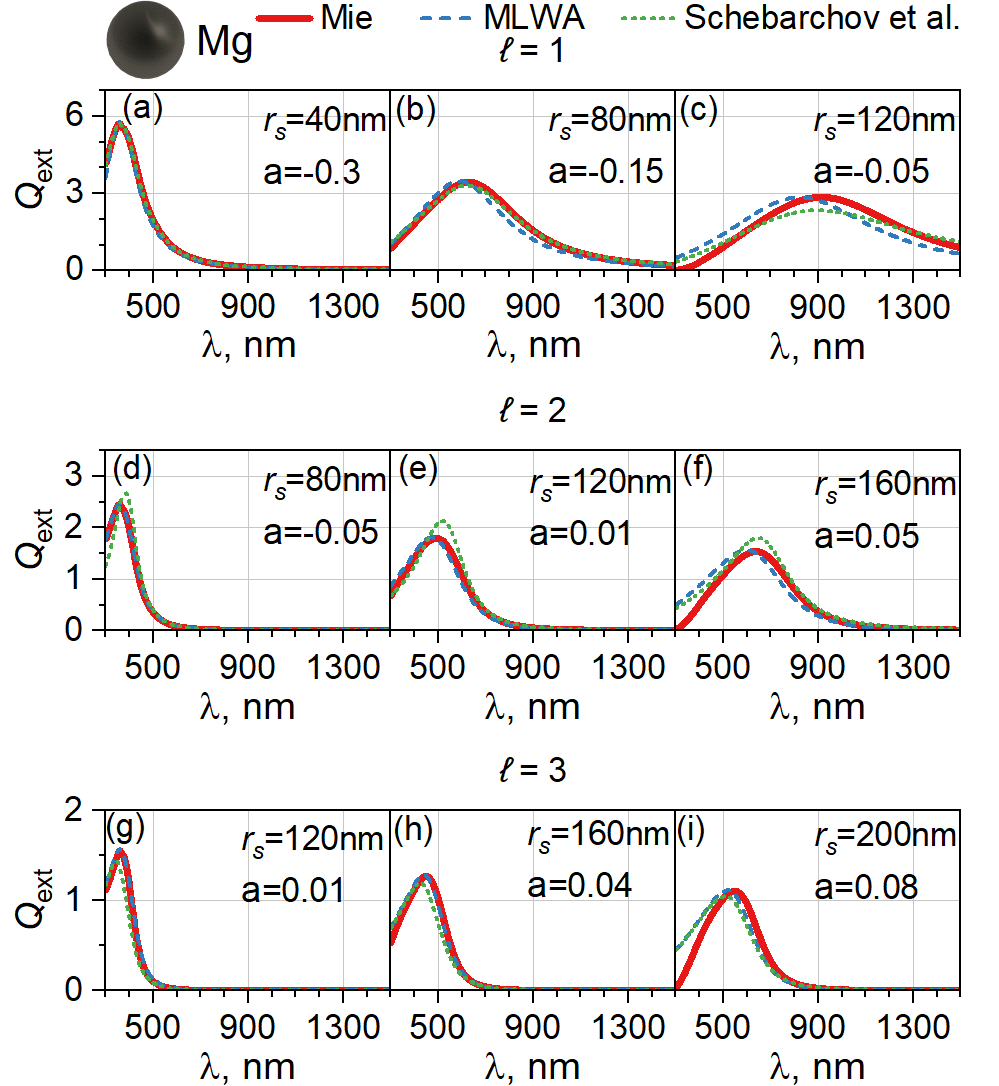}
\caption{The same as in Figure~\ref{fig:Al_occam}, but for Mg spheres.
}
\label{fig:Mg_occam}
\end{figure}

\begin{figure}
 \centering
 \includegraphics{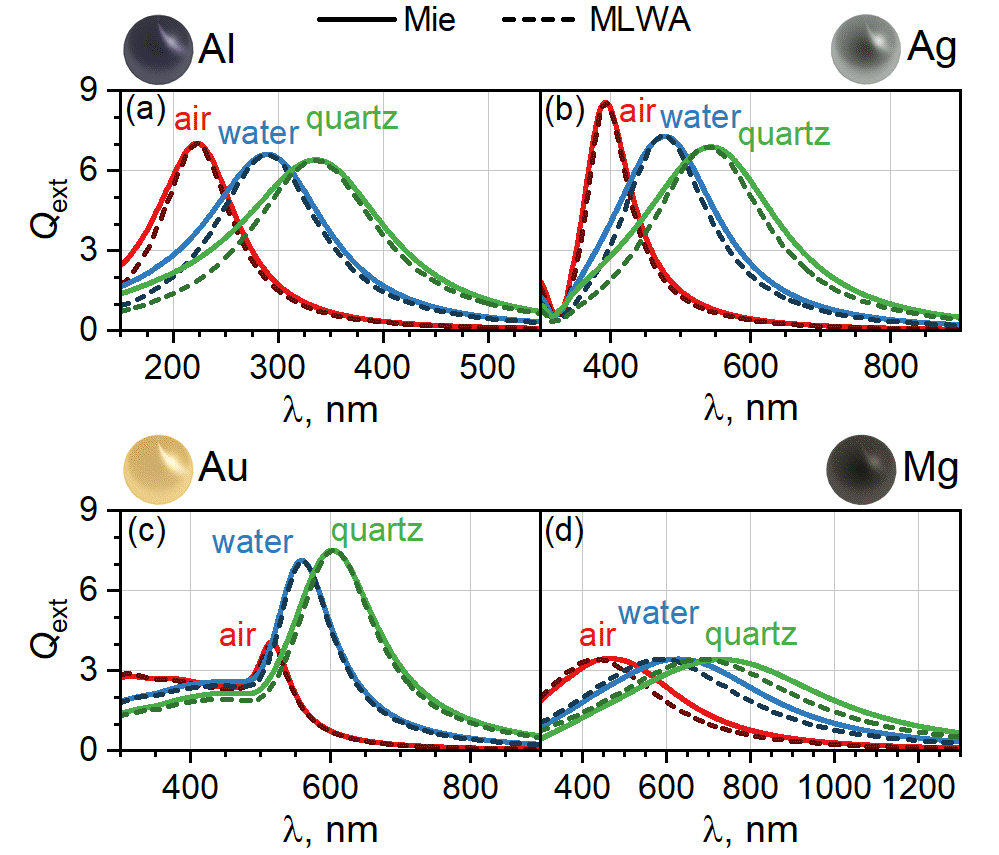}
 \caption{Electric dipole ($\ell = 1$) contribution to the extinction efficiency of Al, Ag, Au, and Mg nanospheres with $r_s=30$~nm, $r_s=50$~nm, $r_s=50$~nm, and $r_s=80$~nm, respectively, embedded in air ($n_h=1.00$), water ($n_h=1.33$) and quartz ($n_h=1.50$) medium.
 Spectra are calculated via exact Mie theory (eq \ref{miecoef}) shown in full line, and via MLWA (eq \ref{mlwaff}) shown by dashed line, with dynamic depolarization from eq~\ref{gDdf} using optimized values of parameter $\mathfrak{a}$ (for air, water, and quartz, respectively): (a)~$-0.29$, $-0.29$, $-0.29$; (b)~$-0.3$, $-0.25$, $-0.23$; (c)~$-0.41$, $-0.37$, $-0.33$; (d) $-0.11$, $-0.13$, $-0.14$.
 }
 \label{fig:host}
\end{figure}

The results shown in Figures \ref{fig:Al_occam}--\ref{fig:Mg_occam} provide a 
clear demonstration of that our ${\cal O}(x^2)$ MLWA surprisingly yields very accurate results for plasmonic spheres both for (i) $x$ up to $\gtrsim 1$ and beyond, and (ii) higher order multipoles ($\ell>1$), essentially {\em doubling} its expected range of validity. The latter is, as shown in Figure~\ref{fig:host}, independent of a host.
Quite unexpectedly, the precision of our results with an optimized $D$ can be noticeably better than that involving the ${\cal O}(x^4)$ approximation of ref~\citenum{Schebarchov2013}, which is shown for a comparison in Figures \ref{fig:Al_occam}--\ref{fig:Mg_occam}.
This observation becomes even more striking after analyzing ${\cal O}(x^4)$ asymptotics ($x\ll 1$) for spherical Bessel functions and their fractions for arbitrary $\ell$ (see Supporting Information) and realizing that its range of validity extends well beyond $x=1$ with deviations less than $1\%$ with respect to exact results (Figures S1-S3, see the Supporting Information).

\newpage

\section{Discussion and Open Questions}
\label{sc:disc}
There are a number of novel features in our approach.
For instance, an ambiguity of the MLWA has been 
spotted earlier~\cite{Zeman1984,Moroz2009,Schebarchov2013}, yet the infinite, one-parameter, continuous family of MLWA's has not been noted before.
Some partial extension of the MLWA for $\ell=2,3$ has been presented by Schebarchov et al.~\ct{Schebarchov2013}, however no general formulas valid for any $\ell$ have been presented. Other points are discussed below.

\subsection{Comparison of ${\cal O}(x^2)$ and ${\cal O}(x^4)$ Approximations}
\label{sc:ox2vsx4}
The MLWA form of $T_{E1}$ (eq \ref{tmtflr3c1}) as discussed 
in ref~\citenum{Moroz2009} (see eq A3 therein) has been improved with a 
quartic $\sim x^4$ term in its denominator (eq 33 of ref~\citenum{Schebarchov2013}),
and its domain of validity has been investigated (Figure 3 of ref~\citenum{Schebarchov2013}), together with its generalization for $\ell=2,3$ (Figure 4 of ref~\citenum{Schebarchov2013}). A distinct advantage of the ${\cal O}(x^4)$ approximation of Schebarchov et al.~\ct{Schebarchov2013} is that one has a fixed formula for all particle parameters. 
However, the ${\cal O}(x^4)$ approximation disguises that our ${\cal O}(x^2)$ MLWA, with an optimized dynamic depolarization term $D$, can actually improve 
on the predictive power of the ${\cal O}(x^4)$ approximation. The very fact that by dropping ${\cal O}(x^4)$ terms and keeping only terms ${\cal O}(x^2)$, while optimizing $D$, one can improve precision, as highlighted in Figures \ref{fig:Al_occam}--\ref{fig:host}, is not only surprising but also bears important physical consequences. Indeed, in a bottom-up approach to depolarization~\ct{Moroz2009}, the absence of any ${\cal O}(x^4)$ term and the sole presence of only ${\cal O}(x^2)$ term in depolarization is a consequence of assuming the internal field $\vE_{in}$ within the particle to be {\em uniform} and {\em constant} on the application of a constant applied field $\vE_0$. The latter is rigorously true in electrostatics for a particle with a general ellipsoid shape. With an increasing particle size, this is less and less true and deviations from the uniformity of $\vE_{in}$ begins to grow. The fact that ${\cal O}(x^2)$ MLWA remains still very good approximation signifies that it is possible to describe the particle properties with a kind of a uniform effective internal field $\bar\vE_{in}$. 
The fitting parameter $\mathfrak{a}$ of $D$ in eq \rf{gDdf} can be seen as allowing us to correct for the averaged internal field $\bar\vE_{in}=\int_V \vE_{in}\, d^3 \vr$, where the integration is over a particle volume.
(Note in passing that for $x\gtrsim 1$, and especially for particle sizes exceeding twice a metal skin depth, we can no longer assume $\vE_0$ to be constant over particle volume).
Because the profile of $\vE_{in}$ is expected to change with particle radius $r_s$, the latter could explain why a given radius subinterval requires its own $\mathfrak{a}$ for the best fit. A dependence of $\mathfrak{a}$ on $r_s$ should not be confused with a size $x$-dependence, because $\mathfrak{a}$ does {\em not} depend on the wavelength $\ld$ (cf. Figures \ref{fig:Al_occam}--\ref{fig:host}).

Last but not the least, recent work by Januar et al.~\cite{Januar2020} has shown that the ${\cal O}(x^2)$ MLWA description of a Drude-like plasmonic particle translates straightforwardly into an equivalent intuitive description of scattering and near-field properties in terms of a driven damped harmonic oscillator 
with the Abraham-Lorenz force, mass, and stiffness all directly related to corresponding partial depolarization terms \cite{Januar2020}. Our work extends the validity of the above intuitive description to much larger particle sizes, whereby
significantly improved understanding of nanoparticle scattering and near-field properties can be achieved.

\subsection{MLWA and Pad\'e Approximation}
\label{sc:pde}
The MLWA form (eq~\ref{mlwaff}) with the size-independent quasi-static Fr\"ohlich term (eq \rf{FT}), a dynamic depolarization term 
$D$ (eq \ref{gDdf}), and the radiative reaction term $R_{E\ell}$ (eq \rf{radcr}) is in its essence a rational polynomial approximation to the exact T-matrix for each given $\ell$. In order to address an intrinsic ambiguity of $D$ term in 
a ``fixed formula for all particle parameters'' approach, a useful criterion could be how a particular $\ell$th channel MLWA compares
against the so-called $[(2\ell+1)/(2\ell+1)]_{T_{p\ell}}(x)$ Pad\'e 
approximation~\cite{Baker2010,Press2007} of the exact T-matrix $T_{p\ell}$.
For this purpose one has to compare the first $4\ell+2$ derivatives 
at $x=0$ of the MLWA ($M$ in eqs SI.21, see Supporting Information) against those of the rigorous $\ell$-channel $T$-matrix ($f$ in eqs SI.21, see Supporting Information). In this regard note 
that any change of $D$ induces changes in the Taylor expansion of a $\ell$th channel MLWA in any order $\sim x^{2\ell+1+2k}$, $k\ge 1$.
If the respective derivatives do agree, then the MLWA can be seen as the Pad\'e approximation of the T-matrix.
One can compute the first $4\ell+2$ derivatives for the functional MLWA form \ref{mlwaff} (see Supplementary Information). 

\section{Summary and Conclusions}
\label{sc:concl}
An intermediate range of sizes ($r_s\gtrsim 20$~nm) of nanoparticles 
clearly shows inadequacy of the Rayleigh limit in their description. 
The lack of unitarity and size dependency, together with further shortcomings, of the latter, 
can be easily overcome by the MLWA,
which provides an economic and concise description of photonic properties of nanoparticles
in any multipole order $\ell$ in terms of a size-independent quasi-static Fr\"ohlich term $F$ (eq~\rf{FT}), a dynamic depolarization term $D$ (eq \ref{gDdf}), 
and a radiative reaction term $R$ ($\sim x^{2\ell+1}$) (eq~\rf{radcr}), all 
combined together in the functional form (eq \ref{mlwaff}).
On making use of that there is an {\em infinite} one parameter set 
of different MLWA's which all satisfy the axioms {\bf A1}-{\bf A2}, 
we have determined for each multipole order $\ell$ an optimal dynamic depolarization term,
which yields the best agreement with the Mie theory. Surprisingly enough, 
such an optimized MLWA has been shown to provide a very reliable description even for particle size parameter 
$x\gtrsim 1$, essentially doubling its expected range of validity, 
which is much larger than has been ever expected to be possible. Our results can be used in a number of different directions and settings:
\begin{itemize}

\item 
Numerical methods such as CDA and DDA~\cite{Purcell1973,Draine1994,Yurkin2007} which divide scatterer 
into a large discrete set of subunits and provide a means for calculating its optical response as the result of interaction of 
elementary dipoles, each corresponding to one of the scatterer subunits, and each described by 
its own dipole polarizability. Since particle dipole polarizability 
$\alpha_1$ is directly related to the dipole $T_{E1}$ by $\alpha_1=-3iT_{E1}/(2k^3)$, 
any method which is intrinsically based on elementary polarizabilities, can, in principle, benefit by adopting our results.

\item 
Using dipole MLWA polarizabilities, one can obtain an intermediary Maxwell-Garnett formula 
improving the usual quasi-static Maxwell-Garnett formula, while providing an insight into the extended Maxwell-Garnett formula~\ct{Ruppin2000,Yannopapas2005}.

\item 
In addition to the dipole MLWA as in ref~\citenum{Moroz2010}, 
the higher-order MLWA results of this paper can be straightforwardly used to further amend the Gersten and Nitzan (GN) quasi-static approximation for determining nonradiative decay rates \cite{Gersten1981} by making use of more precise particle multipolar polarizabilities. Our results could hopefully lead to improving the GN approximation, which at present works for core-shell particles much worse~\cite{Sun2020} than for homogeneous particles~\cite{Moroz2010}. 

\item 
Our precise approximation of $T_{E\ell}$ can be employed also for a deeper understanding of plasmonic sensing~\cite{Otte2010,Jakab2011,Reed2012}, colors~\cite{Wen2016}, and other applications~\cite{Pirzadeh2014} as an interplay of three elementary terms $F$, $D$, and $R$.

\item 
An intuitive description of the dipole contribution of Drude-like plasmonic particles in terms of a driven damped harmonic oscillator~\cite{Januar2020} may offer deeper insight into the above applications.
Keeping only ${\cal O}(x^2)$ terms has a distinct advantage: higher-order ($\ell>1$) MLWA of Drude-like plasmonic particles could, 
in principle, be intuitively described in terms of a driven damped harmonic oscillator with a higher-order 
Abraham-Lorenz force, mass, and stiffness directly related to
corresponding depolarization terms, thereby extending validity of the results 
of ref~\citenum{Januar2020} to much larger particle sizes than initially expected. There is a hope that such an intuitive description could, in addition to a red-shift of near-field
maximum~\cite{Januar2020}, enable to describe also a blue-shift of absorption maximum~\cite{Rasskazov20OL}.

\item Our MLWA expression can be straightforwardly implemented within traditional multiple-scattering theories 
\cite{Korringa1947,Kohn1954,Moroz1995,Moroz1999,Stefanou1998}
to analyze in simple terms the effect of periodic arrangement 
of spherical scatterers in a plane, or in a three-dimensional lattice, 
on various single-sphere multipole contributions.

\item 
Our finding of the one-parameter freedom of the dynamic depolarization term could be of use also for other particle shapes and compositions.
For example, the applicability of the MLWA extends to spheroids~\ct{Zeman1984,Kuwata2003,Moroz2009,Januar2020}, 
which yield for suitable aspect ratios remarkably good approximation for disks (for instance, 
disks with height $20$~nm and radii from $20$ to $250$~nm were approximated as oblate 
spheroids in ref~\citenum{Zoric2011}), rods~\cite{Jakab2011}, cubes 
and cuboids~\cite{Massa2013}, core-shell~\cite{Chung2009,Schebarchov2013}, multilayered~\cite{Chung2010}, and graded-index particles~\cite{Chung2012a}. 

\end{itemize}
Such an economic description of plasmonic properties as an interplay of three elementary terms $F$, $D$, and $R$, provided by the MLWA, is easy to analyze and understand, and thus design nanoparticles with desired properties at the expense of using inherently sophisticated and difficult to understand spherical Bessel and Hankel functions. 
Hopefully in the future no review on plasmonic properties of small metal particles will ever ignore the MLWA~\cite{Fan2014}.

\begin{suppinfo}
${\cal O}(x^4)$ expansion of spherical
Bessel functions and their fractions for arbitrary $\ell$; MLWA derivations; MLWA and driven damped harmonic oscillator model; Pad\'e approximation.
\end{suppinfo}

\bibliography{references}

\end{document}


\usetagform{supplementary}

\section{${\cal O}(x^4)$ Expansion of Spherical
Bessel Functions and Their Fractions for Arbitrary $\ell$}
\lb{sc:ox4}
The results of this section allows us to generalize 
the order ${\cal O}(x^4)$ approximation of Schebarchov et al.~\ct{Schebarchov2013}
for $\ell >3$. In doing so, we shall generalize Lewin's function and its asymptotic~\cite{Lewin1947} for an arbitrary $\ell$.

Asymptotic expansion of the familiar spherical Bessel functions $j_\ell$ and 
$n_\ell$ for $z\rightarrow 0$ involving the 
first three orders is~\cite{Abramowitz1973}:
\begin{eqnarray}
(10.1.2):\, 
j_{\ell}(z) &\sim& 
 \frac{z^\ell}{(2\ell+1)!!} 
 \left[ 1 - \frac{z^2}{2(2\ell+3)} + \frac{z^4}{8(2\ell+3)(2\ell+5)}\right],
\nn\\
{}
 [zj_{\ell}(z)]' &\sim& 
 \frac{z^\ell}{(2\ell+1)!!} \left[ (\ell+1) - \frac{z^2}{2}\fr{\ell+3}{2\ell+3} 
+ \frac{z^4}{8}\fr{\ell+5}{(2\ell+3)(2\ell+5)}\right],
\nn\\
{}
(10.1.3):\, n_{\ell}(z) &\sim& 
 - \frac{(2\ell-1)!!}{z^{\ell+1}}\, 
 \left[ 1 + \frac{z^2}{2(2\ell-1)} + \frac{z^4}{8(2\ell-3)(2\ell-1)}\right],
\nn\\
{}
 [zn_{\ell}(z)]' &\sim& \frac{(2\ell-1)!!}{z^{\ell+1}}\, 
 \left[ \ell + \frac{z^2}{2} \frac{\ell-2}{2\ell-1} 
 + \frac{z^4}{8}\frac{\ell-4}{(2\ell-3)(2\ell-1)}\right],
\lb{spbslimas2}
\end{eqnarray}
where the number in parenthesis on the left corresponds to the 
formula of ref \citenum{Abramowitz1973}. 
Note in passing that although one can use $[zf_{\ell}(z) ]' = zf_{\ell-1}(z)-\ell f_{\ell}(z)$,
which is the recurrence (10.1.21)
of ref \citenum{Abramowitz1973}, to determine $[zf_{\ell}(z) ]'$
for a given spherical Bessel function $f_\ell$, it is much easier
to perform instead direct differentiation of the asymptotic series.
Using eq \rfs{spbslimas2}, one finds
\begin{eqnarray} 
\fr{j_{\ell}(z)}{[zj_{\ell}(z)]'} &\sim & \frac{1}{(\ell+1)}
\left[ 1 + \frac{z^2}{(\ell+1)(2\ell+3)}\right],
\nn\\
\fr{n_{\ell}(z)}{[zn_{\ell}(z)]'} &\sim& -\frac{1}{\ell}
\left[ 1 + \frac{z^2}{\ell(2\ell-1)}\right].
\lb{flgl}
\end{eqnarray}

In order to determine ${\cal O}(x^4)$ correction,
one has to expand
\bea
\fr{1}{ [zj_{\ell}(z)]'} &\sim& \frac{(2\ell+1)!!}{(\ell+1) z^\ell} \left[1 - \frac{(\ell+3) z^2}{2(\ell+1)(2\ell+3)} 
+ \fr{(\ell+5)z^4 }{8(\ell+1)(2\ell+3)(2\ell+5)} \right]^{-1}
\nn\\
 &\sim& \frac{(2\ell+1)!!}{(\ell+1) z^\ell} \left(1 - A \right)^{-1}
\nn\\
 &\sim& \frac{(2\ell+1)!!}{(\ell+1) z^\ell}\left(1 + A + A^2 \right),
 \lb{Fratio}
\eea
and collect in each of $A$ and $A^2$ the terms up to ${\cal O}(z^4)$.
The latter means keeping whole 
\bg
A:= \frac{(\ell+3) z^2}{2(\ell+1)(2\ell+3)} - \fr{(\ell+5)z^4 }{8(\ell+1) (2\ell+3)(2\ell+5)},
\lb{Atrm}
\eg
whereas, in the case of $A^2$, keeping in eq \rfs{Fratio} only the term
\bg
A^2 \sim \frac{(\ell+3)^2 z^4}{4(\ell+1)^2 (2\ell+3)^2}\cdot \nn
\eg
The ${\cal O}(z^4)$-term of the ratio in eq \rfs{Fratio} is thus
\bg
{\cal O}(z^4) = \frac{\ell [\ell (2 \ell+19)+68]+75}{8 (\ell+1)^2 (2 \ell+3)^2 (2 \ell+5)} z^4 \ . \nn
\eg

Using the above results, it is straightforward to determine the asymptotic of
the Bessel functions fractions
\bea
F_\ell(z) &:=& \fr{(\ell+1) j_\ell(z) }{[zj_\ell(z)]' } 
= \fr{(\ell+1) j_\ell(z) }{zj_{\ell-1}(z)-\ell j_\ell(z) },
 \nn\\
G_\ell(z) &:=& - \fr{\ell n_\ell(z) }{[zn_\ell(z)]'} 
= \fr{\ell n_\ell(z) }{\ell n_\ell(z) - zn_{\ell-1}(z) } \cdot
\lb{fmgmd}
\eea
In arriving at the second equalities in \rfs{fmgmd}, 
the recurrence (cf eq 10.1.21 of \citenum{Abramowitz1973})
\bg
[zf_\ell(z) ]' = zf_{\ell-1}(z)-\ell f_\ell(z) \nn
\eg 
satisfied by the spherical Bessel functions has been used.

The ${\cal O}(z^4)$-term of the ratio in the first formula in \rfs{flgl} 
is formally obtained by making use of the formula
\bg
(1+Az^2+Bz^4)(1+Cz^2+Dz^4)\sim 1+(A+C)z^2+(B+D+AC) z^4 +\ldots, \nn
\eg
where the first factor is $j_\ell$ expansion in the 2nd line of eqs \rfs{spbslimas2}, 
and the second factor is the resulting ${\cal O}(z^4)$ expansion in eq \rfs{Fratio}.
In the formula here, $A$ is no longer given by a temporary label in eq \rfs{Atrm} 
but rather ${\cal O}(z^2)$-term of the expansion of $j_\ell$ 
in the 2nd line of eqs \rfs{spbslimas2}.
The ultimate ${\cal O}(z^4)$ term reads as
\bg
{\cal O}(z^4) = \frac{3(2+\ell)z^4}{(5+2\ell)(3+5\ell+2\ell^2)^2}\cdot \nn
\eg
Thus, combining everything together, 
\be
F_\ell(z) \sim 1 + \frac{z^2}{(\ell+1)(2\ell+3)} 
+ \frac{3(\ell+2)}{2\ell+5}\left[ \frac{z^2}{(\ell+1)(2\ell+3)}\right]^2\cdot
\lb{Flzas}
\eg
One can verify that, for $\ell=1$, the familiar asymptotic of Lewin's function $F_1$ is recovered,
\be
F_1(z) \sim 1 + \frac{z^2}{10} + \frac{9z^4}{700}\cdot
\lb{fmxasp}
\eg

Similarly, repeating all the above steps for the second equation in eq~\eqref{flgl}, we get:
\be
G_\ell(z) \sim 1 + \frac{z^2}{\ell(2\ell-1)} 
+ \frac{3(\ell - 1)}{2\ell-3} \left[\frac{z^2}{\ell(2\ell-1)} \right]^2 .
\lb{Glzas}
\eg
Fig \rf{fig:fg} is crucial for understanding both the range of validity of the MLWA and why 
the ${\cal O}(x^2)$ approximation can, in principle, be at least as good as the ${\cal O}(x^4)$ approximation.
One sees that both approximations, in the case of either $F_\ell$ or $G_\ell$, essentially overlie with 
the corresponding exact expression of $F_\ell$ or $G_\ell$ on the real axis.
For instance, each of the ${\cal O}(x^2)$ and the ${\cal O}(x^4)$ approximations \rfs{fmxasp} deviates from the
exact values of $F_1(z)$ for {\em real} $x$ with not more than 1\%
for $x$ up to $x\approx 1.34$. The respective ${\cal O}(x^2)$ and ${\cal O}(x^4)$ approximations
begin to significantly deviate
with the corresponding exact expression of $F_\ell$ or $G_\ell$ at nearly the same point of the real axis.
\begin{figure}
\centering
\includegraphics[width=3.33in]{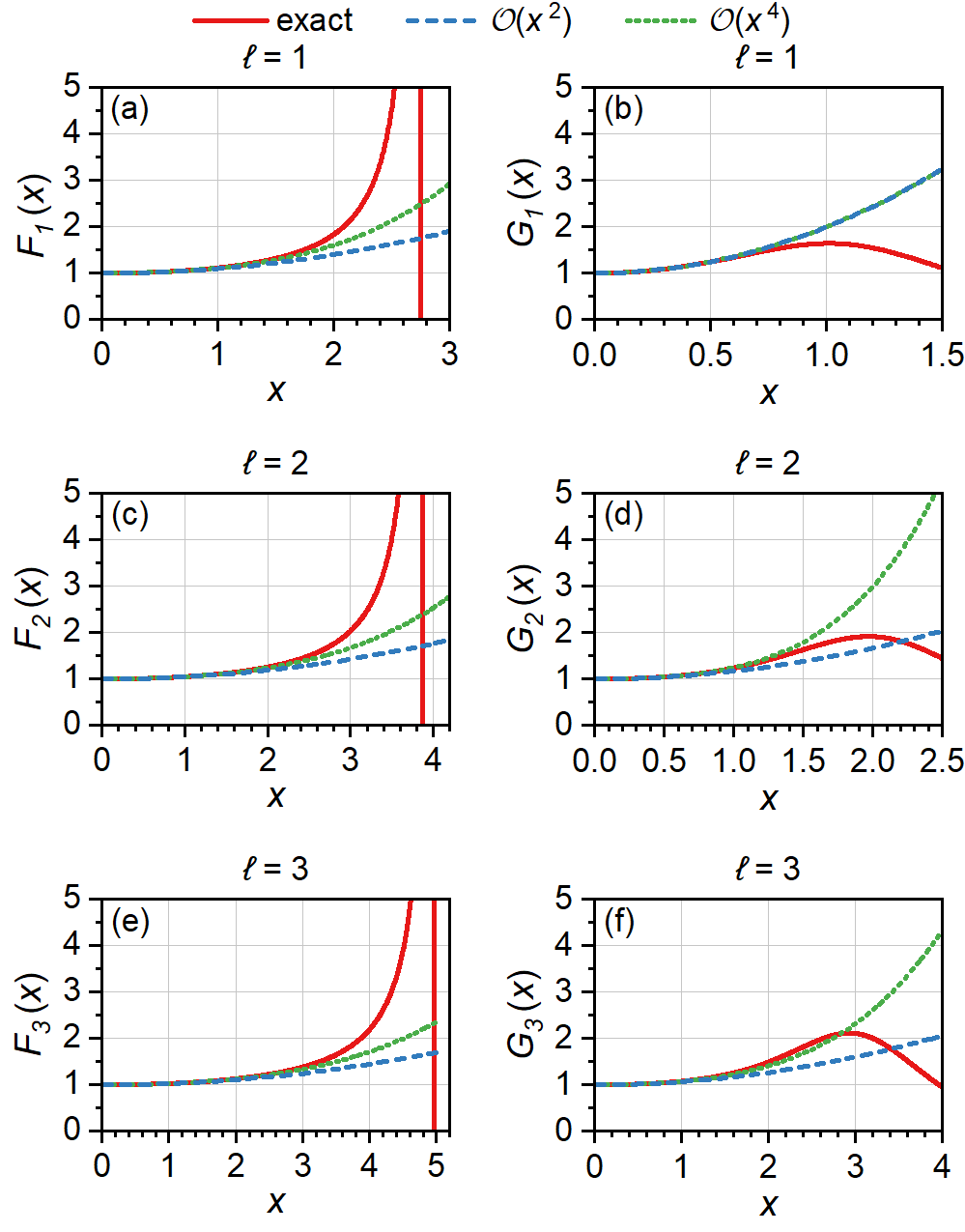}
\caption{A graphical illustration of the range of validity 
of the respective ${\cal O}(x^2)$ and ${\cal O}(x^4)$ asymptotic expansions of $F_\ell$ and $G_\ell$ showing that they can be approximated on the real axis with the accuracy higher than $1\%$ for $\ell\le 3$. Note is passing that $F_\ell$ can be approximated
on a larger interval of the real axis well beyond $x\gtrsim 1$ than corresponding $G_\ell$. However, any attempt to approximate $F_\ell$ on a real interval any further eventually breaks down due to the first pole of $F_\ell$.
}
\label{fig:fg}
\end{figure}

\begin{table}
 \caption{Values of $x_{max}$ for which ${\cal O}(x^2)$ and ${\cal O}(x^4)$ approximations of $F_\ell(x)$ and $G_\ell(x)$ given in eqs~\rfs{Flzas} and \rfs{Glzas} are valid with up to 1\% accuracy for $0<x<x_{max}$ (cf. Figure~\ref{fig:fg})}
 \label{tbl:o2o4}
 \begin{tabular}{l|l|l|l|l|l}
 \hline
 \multicolumn{3}{c|}{$F_\ell(x)$} & \multicolumn{3}{|c}{$G_\ell(x)$} \\
 \hline
 $\ell$ & ${\cal O}(x^2)$ & ${\cal O}(x^4)$ & $\ell$ & ${\cal O}(x^2)$ & ${\cal O}(x^4)$\\
 \hline
 1 & 0.92 & 1.34 & 1 & 0.57 & 0.57 \\
 2 & 1.34 & 1.89 & 2 & 0.61 & 0.99 \\
 3 & 1.74 & 2.46 & 3 & 0.99 & 1.34
 \end{tabular}
\end{table}

\begin{figure}[t!]
\centering
\includegraphics[width=3.33in]{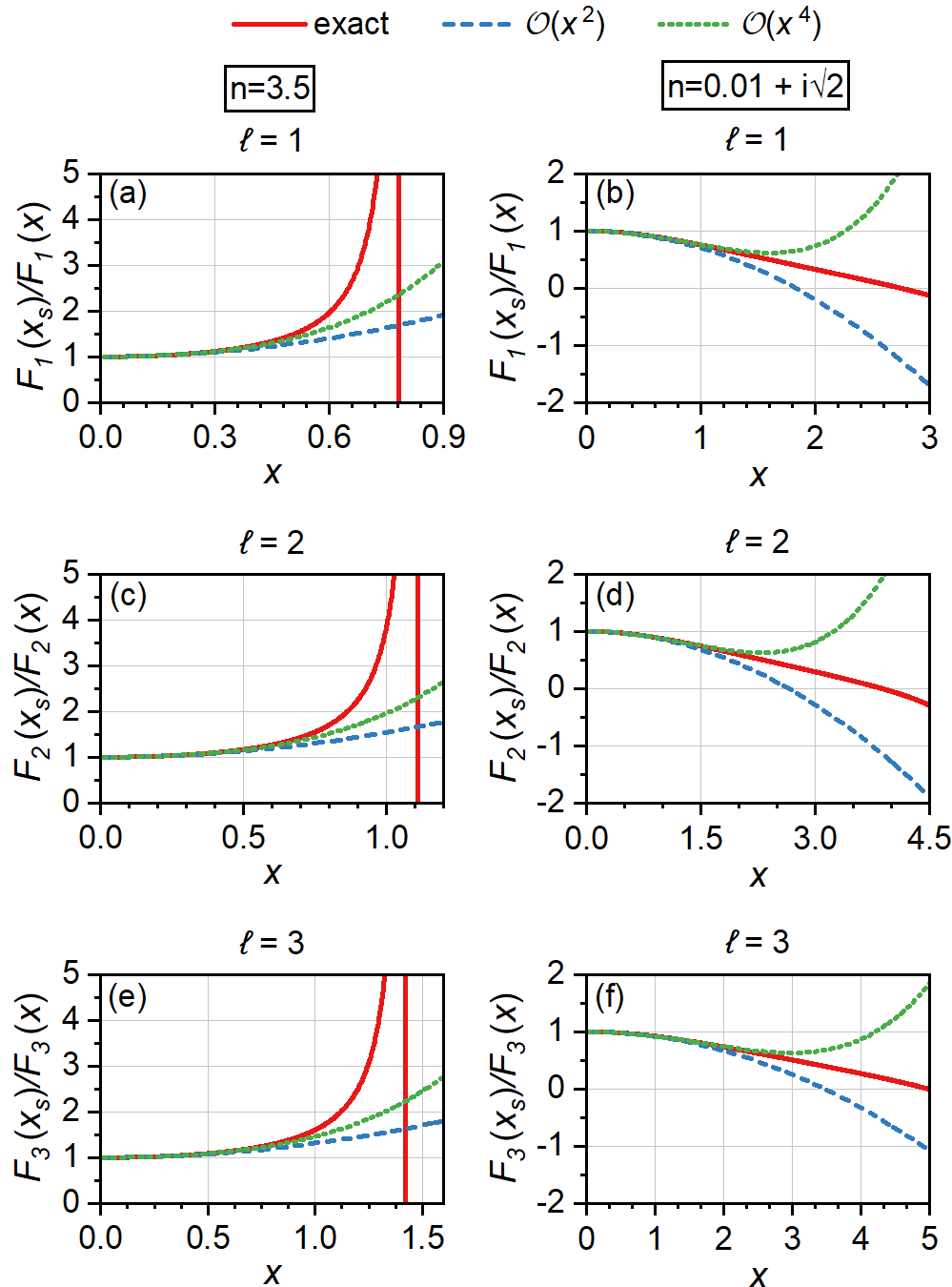}
\caption{A graphical illustration of the range of validity 
of the respective ${\cal O}(x^2)$ and ${\cal O}(x^4)$ asymptotic expansions of $F_\ell (x_s)/F_\ell (x)$ for real (left) and complex (right) relative refractive index contrast, $n=n_s/n_h$. The right panels here corresponds to $\veps=n^2\approx -2$, i.e. to a proximity of the dipole LSPR. 
Similarly to Figure \ref{fig:fg}, ${\cal O}(x^2)$ and ${\cal O}(x^4)$ approximations essentially overlie with 
the corresponding exact expression for a long stretch of $x$ on the real axis.
}
\label{fig:ff_cr}
\end{figure}

\begin{figure}[t!]
\centering
\includegraphics[width=3.33in]{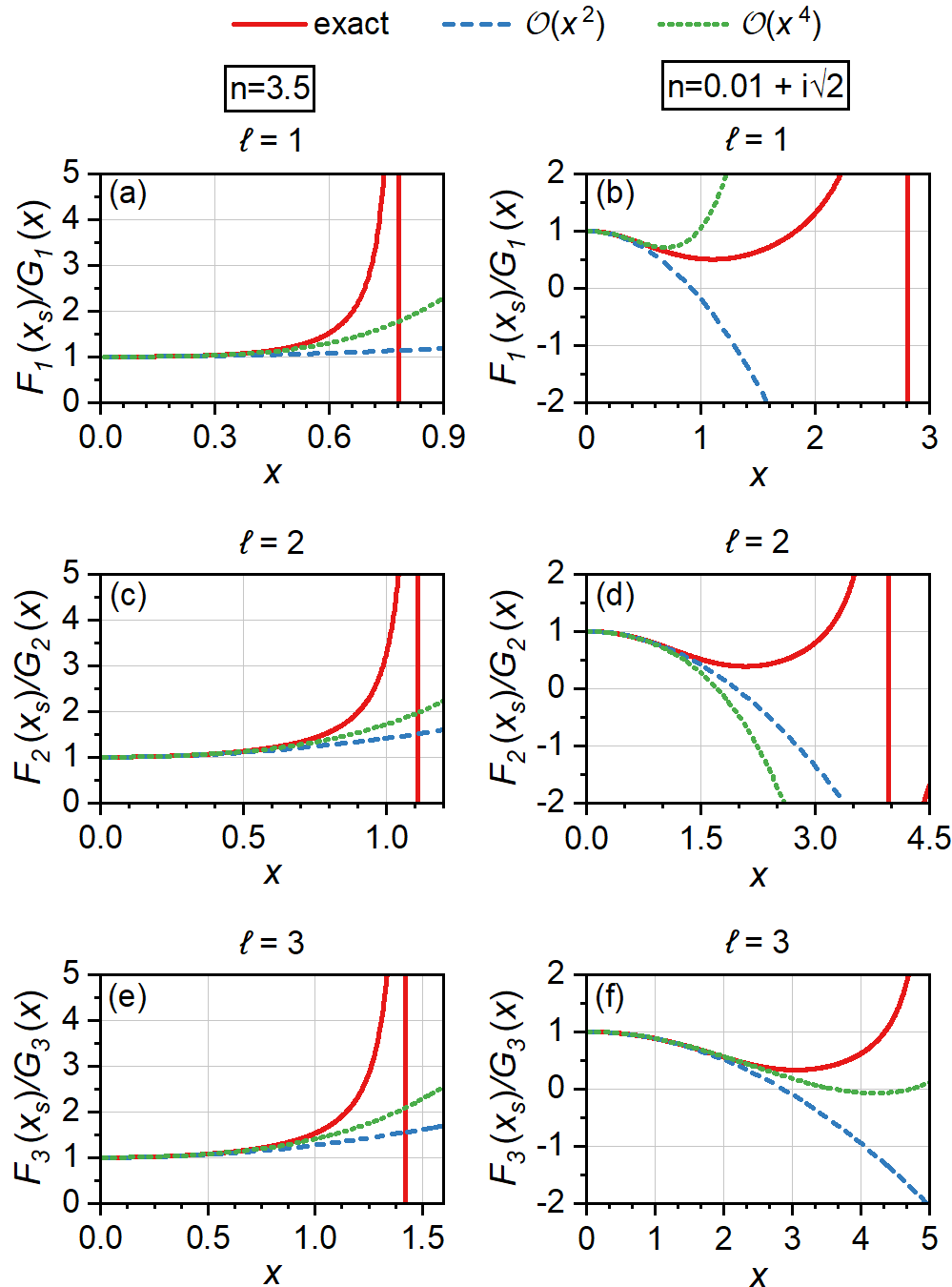}
\caption{The same as in Figure~\ref{fig:ff_cr}, but for $F_\ell (x_s)/G_\ell (x)$. 
Crucially, and much the same in Fig. \rf{fig:ff_cr}, the respective ${\cal O}(x^2)$ and ${\cal O}(x^4)$ approximations overlie in a proximity of the dipole LSPR for $\veps=n^2\approx -2$, as shown in the right panels, over longer $x$ interval with the exact expressions. This is the reason why MLWA works better for plasmonic particles than for purely dielectric particles.
}
\label{fig:fg_cr}
\end{figure}

It has been known that, for purely {\em real} values of $x$ in the dipole case,
$F_1(x)$ is monotonically increasing with increasing $x$
up to the first pole of $F_1(x)$ at $x=2.74370725179062$. We see analogous behaviour also for $\ell>1$,
with the pole position shifting to larger $x$ with increasing $\ell$.

On following similar steps as above, one obtains the following asymptotic of the ratios
\bea
\lefteqn{
F_\ell (x_s)/F_\ell (x) \sim 1 + \fr{(n^2-1) x^2}{(\ell+1)(2\ell+3)} 
}
\nn\\ 
&&
+ \, \left[ \frac{3(\ell+2) (n^4 - 1) - (2\ell+5)n^2}{2 \ell+5}+\frac{1}{(\ell+1)^2 (2 \ell+3)^2}\right] \left[\frac{x^2}{(\ell+1)(2\ell+3)}\right]^2,
\nn\\
\lefteqn{
F_\ell (x_s)/G_\ell (x) \sim 1 + 
 \left[\fr{n^2}{(\ell+1)(2\ell+3)} -\fr{1}{\ell(2\ell-1)} \right] \, x^2}
 \nn\\
 &&
 +\, \left[ \frac{3(\ell+2)n^4}{(2\ell+5)(\ell+1)^2(2\ell+3)^2} - \frac{n^2}{\ell(\ell+1)(2\ell-1)(2\ell+3)} - \frac{1}{\ell(2\ell-3)(2\ell-1)^2} \right]x^4.
 \nn\\
\lb{flgll4}
\eea
The above ratios play a very significant role
in determining the so-called {\em renormalized} 
dielectric permittivity and magnetic permeability as in eq \rfs{plpl} below.

\begin{table}[h]
 \caption{Values of $x_{max}$ for which ${\cal O}(x^2)$ and ${\cal O}(x^4)$ approximations of $F_\ell(x_s)/F_\ell(x)$ and $G_\ell(x_s)/F_\ell(x)$ given in eqs~\rfs{flgll4} are valid with up to 1\% accuracy for $0<x<x_{max}$ (cf. Figures~\ref{fig:ff_cr} and \ref{fig:fg_cr}). The complex value of $n$ corresponds to $\veps=n^2\approx -2$, i.e. to a proximity of the dipole LSPR.}
 \begin{tabular}{l|l|l|l|l|l|l|l|l|l|l|l}
 \multicolumn{6}{c|}{$n=3.5$} & \multicolumn{6}{|c}{$n=0.01+i\sqrt{2}$} \\
 \hline
 \multicolumn{3}{c|}{$F_\ell(x_s)/F_\ell(x)$} & \multicolumn{3}{|c|}{$F_\ell(x_s)/G_\ell(x)$} &
 \multicolumn{3}{|c|}{$F_\ell(x_s)/F_\ell(x)$} & \multicolumn{3}{|c}{$F_\ell(x_s)/G_\ell(x)$} \\
 \hline
 $\ell$ & ${\cal O}(x^2)$ & ${\cal O}(x^4)$ & $\ell$ & ${\cal O}(x^2)$ & ${\cal O}(x^4)$ & 
 $\ell$ & ${\cal O}(x^2)$ & ${\cal O}(x^4)$ & $\ell$ & ${\cal O}(x^2)$ & ${\cal O}(x^4)$\\
 \hline
 1 & 0.27 & 0.38 & 1 & 0.27 & 0.40 & 1 & 0.62 & 1.06 & 1 & 0.30 & 0.46 \\
 2 & 0.39 & 0.55 & 2 & 0.42 & 0.56 & 2 & 0.89 & 1.51 & 2 & 1.08 & 0.80 \\
 3 & 0.50 & 0.71 & 3 & 0.52 & 0.72 & 3 & 1.16 & 1.95 & 3 & 1.43 & 1.47
 \end{tabular}
 \label{Tb1}
\end{table}

\vst{0.4cm}

The precision of the above asymptotic formulas, and their range of applicability,
strongly depend on how $x_s$ moves in the complex plane of $x_s$ with varying $\ld$. As seen in Table \rf{Tb1} and Figures~\ref{fig:ff_cr} and \ref{fig:fg_cr}, the real axis provides the largest restriction on the use 
of above asymptotic formulas. Contrary to that, in a Drude-like regime, the complex
refractive index $n=n_r+i\kappa$, $\kappa\ge 0$ is necessarily 
dominated in magnitude by its imaginary part $\kappa$. The latter is obvious
from that $n^2 =n_r^2-\kappa^2 +2in_r\kappa$ has to reproduce the Drude dielectric function 
characterized by a negative real part, which magnitude can be appreciable.
Therefore, for metal particles, $x_s$ moves in the complex plane of $x_s$ closer to the 
positive imaginary axis than to the real axis, as is the case of purely dielectric particles.
The latter is the main reason of a much larger range validity of the MLWA for plasmonic particles compared
to dielectric ones, as demonstrated in Table \rf{Tb1} and Figures~\ref{fig:ff_cr} and \ref{fig:fg_cr}.

\section{MLWA Derivations}
\lb{sc:qsl}
In this section, we derive a general form of transfer-matrix $T_{E\ell}$ in the MLWA limit and arrive to eqs 9--12. In doing so we also generalize 
the order ${\cal O}(x^4)$ approximation of Schebarchov et al.~\ct{Schebarchov2013} for an arbitrary $\ell$.
The asymptotic expansions are used in accordance with our rules ({\bf R1})-({\bf R2}). In order to make it unambiguous, we will be expanding spherical Bessel functions in each of the numerator and denominator of the fraction of eq 4 defining the $T$-matrix,
and {\em not} expanding the fraction of eq 4 into the Taylor series at $x=0$.

Scattering can be equally well described by any of 
the three familiar inter-related $T$, $S$, and $K$ matrices:
\be
T = \fr{1}{2} \, (S-1), \qquad S=1+2T, \qquad
K = \dfrac{iT}{1+T} = - i \dfrac{1-S}{1+S}, \qquad S= \dfrac{1-iK}{1+iK}, \nn
\eg
where the $S$ matrix is related by the Cayley transform with the inverse $K^{-1}$ of the $K$ matrix.
This is not surprising, because the Cayley transform $f(z)=(z-i)/(z+i)$ maps the real 
line to the unit circle, and the unitarity of the $S$ matrix amounts to reality 
of the $K$ matrix. In terms of a scattering phase shift $\eta$, one has $S=e^{2i\eta}$, 
$T=i\sin\eta e^{i\eta}$, and $K=-\tan \eta$.

In the present 3D problem of electromagnetic scattering from a sphere,
\bea
K_{p\ell} &=& - \fr{j_\ell(x)[x_sj_\ell(x_s)]' - \upsilon j_\ell(x_s) [xj_\ell(x)]' }
{n_\ell(x)[x_sj_\ell(x_s)]' - \upsilon j_\ell(x_s) [xn_\ell(x)]' }
\nn\\
&=& \fr{\upsilon j_\ell(x_s) [xj_\ell(x)]' - j_\ell(x)[x_sj_\ell(x_s)]' }
{- \upsilon j_\ell(x_s) [xn_\ell(x)]' + n_\ell(x)[x_sj_\ell(x_s)]' }
\nn\\
&=& \fr{j_\ell(x)}{n_\ell(x)} 
\fr{\upsilon F_\ell(x_s)/F_\ell(x) - 1 }{\upsilon [F_\ell(x_s)/G_\ell(x)] [\ell/(\ell+1)] +1 }, \nn
\eea
where $\upsilon=\mu_s/\mu_h$ for $p=M$, and $\upsilon=\veps_s/\veps_h$ for $p=E$, 
$j_\ell$, $n_\ell$ are the usual spherical Bessel functions (see Sec. 10 of ref~\citenum{Abramowitz1973}),
and the shorthands $F_\ell$ and $G_\ell(x)$ have been introduced by eq \rfs{fmgmd}. Let
\bg
\tl{\upsilon}_\ell = \upsilon F_\ell(x_s)/F_\ell(x), \hst{1.5cm}
\hat{\upsilon}_\ell = \upsilon F_\ell(x_s)/G_\ell(x).
\lb{plpl}
\eg
Then
\bg
K_{p\ell} = \fr{j_\ell(x)}{n_\ell(x)}\, \fr{\tl{\upsilon} - 1 }{\ell \hat{\upsilon}/(\ell+1) +1 }\cdot
\lb{Kell}
\eg
The function $F_\ell(x_s)$ has been for $\ell=1$ 
introduced by Lewin \cite{Lewin1947} and 
employed by e.g. Sarychev, McPhedran, and Shalaev \cite{Sarychev2000,*Sarychev2001} 
and many others \ct{Yannopapas2005} to define, as in the last two equations above, the so-called {\em renormalized} 
dielectric permittivity and magnetic permeability. 

The advantage of making use of the $K$ matrix is that it is {\em real} 
in purely dielectric case and enables one to write the $T$ matrix as
\bg
T_{p\ell} = \frac{-iK_{p\ell}}{1+iK_{p\ell}}\cdot
\lb{tAkA}
\eg
Now, the first fraction on the rhs of eq \rfs{Kell} reduces for $x\ll 1$ to a numerical 
prefactor $-(\ell+1) x^{2\ell+1}/(2\ell-1)!! (2\ell+1)!!$ (see formulas 10.53.1-2 in ref~\citenum{Olver2010}), 
which enables one to introduce a renormalized polarization factor
\bg
Q_{p\ell} = - \fr{(2\ell-1)!! (2\ell+1)!! }{(\ell+1) x^{2\ell+1} }\, 
K_{p\ell}\to \fr{\tl{\upsilon}_\ell - 1}{\ell \hat{\upsilon}_\ell + (\ell+1) }\qquad (x\ll 1). \nn
\eg
Correspondingly, $T_{p\ell}$ in eq \rfs{tAkA} can be recast as
\bea
T_{p\ell} & =& i\fr{(\ell+1) x^{2\ell+1} }{(2\ell-1)!! (2\ell+1)!! } Q_{p\ell} 
 \left(1 - i\fr{(\ell+1) x^{2\ell+1} }{(2\ell-1)!! (2\ell+1)!! } Q_{p\ell} \right)^{-1}
 \nn\\
 & =& i\fr{(\ell+1)x^{2\ell+1} }{\ell (2\ell-1)!! (2\ell+1)!! } (\tl{\upsilon}_\ell - 1) 
 \left( \fr{\ell+1}{\ell} + \hat{\upsilon}_\ell 
 - i\fr{(\ell+1) x^{2\ell+1} }{\ell(2\ell-1)!! (2\ell+1)!! } (\tl{\upsilon}_\ell - 1) \right)^{-1}.
\nn\\
\lb{tmtf}
\eea
The equation is starting point at arriving at general $\ell$-pole MLWA, eqs 9-12.
The latter requires to consider intermediate values of $x$, which necessitates to keep ${\cal O}(x^2)$ terms in the asymptotic of the ratios
in eqs \rfs{flgll4}, or
\bea
F_\ell (x_s)/F_\ell (x) &\sim& 1 + \fr{(n^2-1)x^2}{(\ell+1)(2\ell+3)},
\nn\\
F_\ell (x_s)/G_\ell (x) &\sim& 1 + 
 \left(\fr{n^2}{(\ell+1)(2\ell+3)} -\fr{1}{\ell(2\ell-1)} \right) \, x^2.
\lb{flgll}
\eea
For $p=E$ polarization, one arrives on combining \rfs{tmtf} and \rfs{flgll} at 
the MLWA limit 
\bea
T_{E\ell} &\sim& 
\fr{ iR (x)}
 { \varepsilon+\fr{\ell+1}{\ell} + \fr{\varepsilon}{(\ell+1)(2\ell+3)} 
 \left(n^2 -\fr{(\ell+1)(2\ell+3)}{\ell(2\ell-1)} \right) 
 \, x^2 - iR (x)},
\lb{tmtfl}
\eea
where
\bg
R(x)=\fr{(\ell+1) x^{2\ell+1} }{\ell(2\ell-1)!! (2\ell+1)!! } 
\left[\varepsilon -1 + \fr{(n^2-1)\varepsilon x^2}{(\ell+1)(2\ell+3)}\right]. \nn
\eg
On substituting $n^2=\varepsilon$ one arrives at our expression 9.
The {\em dynamic depolarization} term $D$ ($\sim x^2$) originates on expanding 
$x$-dependent $\hat{\upsilon}_\ell$ in \rfs{tmtf}.
Obviously, one finds the usual quasi-static limit for $p=E$ upon neglecting the ${\cal O}(x^2)$ terms
in eq \rfs{tmtfl}, or when
\bg
Q_{E\ell} \to \fr{(\veps_s/\veps_h) - 1}
 {\ell (\veps_s/\veps_h) + (\ell+1) }
 =\fr{1}{\ell} \fr{(\veps_s/\veps_h) - 1}
 {\fr{\ell+1}{\ell} + (\veps_s/\veps_h)}\cdot \nn
\eg
is substituted into first equality in eq \rfs{tmtf}.

Note in passing that the form of the limit expression
\rfs{tmtfl} is ambiguous and, by multiplying both the numerator and denominator 
of \rfs{tmtfl} by $1-\varepsilon x^2/[(\ell+1)(2\ell+3)]$, 
it can be reduced to our expression 10,
\bg
T_{E\ell} \sim \fr{iR_{E\ell}}
 {
 \varepsilon + \fr{\ell+1}{\ell} - \fr{2(2\ell+1)}{\ell(2\ell-1)(2\ell+3)} \, \varepsilon x^2 
-iR_{E\ell}}\cdot \nn
\eg

Upon using the asymptotic expansions (Ch. 10 of ref~\citenum{Olver2010}), 
while following our recipe ({\bf R1}) of keeping only terms up to ${\cal O}(x^2)$ 
in the asymptotic expansion of each spherical Bessel function, and that
\bg
\fr{d}{dr} [r f(kr)] = \fr{d}{d (kr) } [ kr f(kr)], \nn
\eg 
the denominator in the expression 4 of the $T$-matrix is given by
\bea
\lefteqn{i [\upsilon v_\ell'(k_2 r_s) u_\ell(k_1 r_s) 
 - v_\ell(k_2 r_s)u_\ell'(k_1 r_s)]=ir_s\, \fr{(2\ell-1)!!}{(2\ell+1)!!}\fr{x_s^\ell}{x^{\ell+1}}\times
}
\nn\\
&&
\left[
\upsilon \left(\ell+\fr{x^2}{2} \fr{\ell-2}{2\ell-1}-\fr{x_s^2}{2} \fr{\ell}{2\ell+3}
\right)+ \ell+1+\fr{x^2}{2} \fr{\ell+1}{2\ell-1}-\fr{x_s^2}{2} \fr{\ell+3}{2\ell+3}
\right],
\nn
\eea
where $u_\ell(x)=xj_\ell(x)$ and $v_\ell(x)=xn_\ell(x)$ are the usual Riccati-Bessel functions (see Sec. 10.3 \citenum{Abramowitz1973}).
Upon taking into account that $x_s^2 = x^2 n_s^2/n_h^2$, 
one finds eventually for the denominator of the $T$-matrix
\bea
\lefteqn{i [\upsilon v_\ell'(k_2 r_s) u_\ell(k_1 r_s) 
 - v_\ell(k_2 r_s)u_\ell'(k_1 r_s)]=ir_s\,\fr{(2\ell-1)!!}{(2\ell+1)!!}\fr{x_s^\ell}{x^{\ell+1}}\times
}
\nn\\
&&
\left\{
\ell \upsilon + (\ell+1) + \fr{x^2}{2(2\ell+3)} 
\left[-\ell \upsilon^2 - \fr{3(2\ell+1)}{(2\ell-1)}\, \upsilon + \fr{(\ell+1)(2\ell+3)}{2\ell-1}
\right] + {\cal O}(x^4) \right\}.
\nn\\
\lb{tdenom}
\eea
The numerator in the expression 4 of the $T$-matrix is given by
\bea
\lefteqn{
[\upsilon u_\ell'(k_2 r_s) u_\ell(k_1 r_s) 
 - u_\ell(k_2 r_s)u_\ell'(k_1 r_s)]
= 
r_s [\upsilon u_\ell'(k_2 r_s) j_\ell(k_1 r_s) 
 - j_\ell(k_2 r_s)u_\ell'(k_1 r_s)] = }
\nn\\
 && r_s \fr{x_s^\ell x^\ell}{[(2\ell+1)!!]^2} \times \,
\nn\\
&&
\left\{
\upsilon \left[ (\ell+1) - \fr{x^2}{2} \fr{\ell+3}{2\ell+3} \right] 
 \left(1 -\fr{x_s^2}{2(2\ell+3)}
\right) - \left(1 -\fr{x^2}{2(2\ell+3)}
\right) \left[ (\ell+1) - \fr{x_s^2}{2} \fr{\ell+3}{2\ell+3} \right] 
\right\}
\nn\\
 &&
= r_s \fr{x_s^\ell x^\ell}{[(2\ell+1)!!]^2}\, (\ell+1) (\upsilon-1) 
\left\{ 1 - (\upsilon+1)\, \fr{x^2}{2(2\ell+3)}
+ {\cal O}(x^4) \right\}.
\lb{tnum}
\eea
Hence, upon combining the numerator asymptotic $N$ given by \rfs{tnum}, and 
the denominator asymptotic $D$ given by \rfs{tdenom},
one finds our MLWA expression 11,
\bea
T_{E\ell} &=& -\fr{N}{N+D}\sim -\fr{N}{D} 
\nn\\
&\sim & 
\fr{i \ell R_{E\ell} \left[1-(\veps+1)\,\fr{x^2}{2(2\ell+3)}\right]}
{
\ell\veps +\ell+1 
+\fr{x^2}{2(2\ell+3)}\left[-\ell\veps^2 -\fr{3(2\ell+1)}{2\ell-1}\,\veps
+\fr{(\ell+1)(2\ell+3)}{2\ell-1}\right]
- i\ell R_{E\ell}
}\cdot
\lb{tmt3o}
\eea
As it has been alluded to in connection with the limit expression
\rfs{tmtfl}, the limit form \rfs{tmt3o} of $T_{El}$ is ambiguous and can, 
by multiplying both the numerator and denominator by $1 + (\veps+1) x^2/[2(2\ell+3)]$, 
be alternatively recast as
\bg
T_{E\ell} \sim \fr{iR_{E\ell}}
 {
 \varepsilon + \fr{\ell+1}{\ell} + 
\left[(\ell-2) \varepsilon + \ell+1\right] \fr{(2\ell+1) x^2}{\ell(2\ell-1)(2\ell+3)} 
 -iR_{E\ell}}, \nn
\eg
which is the MLWA with $D(x)$ of eq 12.

On substituting full formulas \rfs{flgll4} into eq \rfs{plpl}
determining the renormalized
dielectric permittivity and magnetic permeability, one immediately obtains an extension of the ${\cal O}(x^4)$ approximation of Schebarchov et al.~\ct{Schebarchov2013} for an arbitrary $\ell$.
In this case, our rules ({\bf R1})-({\bf R2}) are obviously amended to include also the terms ${\cal O}(x^4)$ in the corresponding asymptotic expansions of Bessel functions.

\section{MLWA and Driven Damped Harmonic Oscillator Model}
\lb{sc:jnr}
In this section, we provide guidelines for using our MLWA (eq 6), with the size-independent quasi-static Fr\"ohlich term (eq 7), a dynamic depolarization term 
$D$ (eq 20), and the radiative reaction term $R_{E\ell}$ (eq 13) in harmonic oscillator model~\cite{Januar2020}.

A free electron of mass $m_e$ and charge $-e$ in an external harmonic field $\vE_0$ 
obeys a simple equation for a driven damped harmonic oscillator, 
\bg
m_e \ddot \vX + m_e\gm \dot \vX = -e\vE_0.
\lb{feho}
\eg
Here $\vX$ denotes the harmonic displacement 
of the electron ($\vX(\om) = \vX_0e^{-i\om t}$) and an overdot stands for time derivative.
Equation \rfs{feho} is a starting point at arriving at the Drude formula for bulk electrons~\cite{Kittel2004,Jackson1999}.
The Drude loss term, $\gm$, is similar to {\em viscous damping}, indicating damping 
that is proportional to the collision rate of electrons, also known as intraband transition damping.

However, within a particle the incoming electric field $\vE_0$ is modified due to the induced depolarization field to $\vE_{in}:=\vE_0 +\vE_d$, where the induced depolarization field,
\bg
\vE_d=- \fr{1}{\varepsilon_h} \vL_{e\!f\!f}\cdot\vP, \nn
\eg 
takes into account the change of external harmonic field $\vE_0$ as 
felt by electrons inside the particle. 
The field is directly proportional to the 
effective depolarization factor $\vL_{e\!f\!f}$ and the polarization $\vP$. 

For our MLWA one finds in the dipole case
\bg
L_{e\!f\!f}(x) = L- \fr{\mathfrak{a}\veps +\mathfrak{b}}{3(\varepsilon-1)}\, x^2 - i\fr{2}{9}\,x^3, \nn
\eg
with $L$ being the well-known {\em geometrical factor} (cf. eq 5.32 of ref \citenum{Bohren1998}), 
which accounts for the shape of a particle. In particular, $L=1/3$ for a sphere.
The dipole polarizability can be recast as
\bg
\al_{\tiny MWLA} = \fr{V}{4\pi} \,\fr{\veps-1}{1 + L_{e\!f\!f}(x) (\veps-1)}\cdot \nn
\eg
To comply with ref~\citenum{Januar2020}, we recast $L_{e\!f\!f}(x)$ as $L_{e\!f\!f}(\om)$
on making use of that dimensionless $x=kr_s=(\om/c) n_h r_s$,
\bg
L_{e\!f\!f}(\om) = L- \fr{\mathfrak{a}\veps +\mathfrak{b}}{3(\varepsilon-1)c^2}\, \om^2 - i\fr{2}{9c^3 }\,\om^3= L-L_d \om^2-i L_{rad} \om^3.
\lb{leffic}
\eg
On using the equation above instead of eqs 9 and 13 of ref~\citenum{Januar2020}, the range of validity of results presented in ref~\citenum{Januar2020} can, in principle, be extended to larger particles, even though $L_{e\!f\!f}(\om)$ in eq \rfs{leffic} is $\varepsilon$-dependent.

\section{Pad\'e Approximation}
\lb{sc:pade}
Given a function $f$ and two integers $s\ge 0$ and $t\ge 1$, 
the Pad\'e approximant of order $[s/t]$ is the rational function
\bg
M(x)=\frac{\sum_{j=0}^s a_j x^j}{1+\sum_{k=1}^t b_k x^k}
=\frac{a_0 +a_1 x+a_2 x^2 +\dots +a_s x^s}{1+b_1 x+b_2 x^2 +\dots +b_t x^t}, \nn
\eg
which agrees with $f(x)$ to the highest possible order $x^{s+t}$, which amounts to
\begin{align}
f(0)&=M(0),
\nn\\
f'(0)&=M'(0),
\nn\\
f''(0)&=M''(0),
\nn\\
&\vdots 
\nn\\
f^{(s+t)}(0)&=M^{(s+t)}(0).
\lb{Pdr}
\end{align}
The Pad\'e approximant defined above is also denoted as $[s/t]_{f}(x)$.

Now consider an explicit form of the MLWA of eq 20, 
\bg
T_{p\ell} \sim 
\fr{iR_{E\ell}(x)}
 {
 \upsilon+ \fr{\ell+1}{\ell} + (\mathfrak{a}\veps +\mathfrak{b}) x^2
 -iR_{E\ell}(x)}\cdot \nn
\eg
%
It is tempting to compare the $\ell$th channel MLWA against 
the so-called $[(2\ell+1)/(2\ell+1)]_{T_\ell}(x)$ 
rational Pad\'e approximation of the corresponding exact $\ell$th channel $T$-matrix $T_{E\ell}$.
In doing so, one has to demonstrate that our MLWA ($M$ in eqs \rfs{Pdr}) 
has identical first $4\ell+2$ derivatives 
at $x=0$ as the rigorous $\ell$-channel $T$-matrix ($f$ in eqs \rfs{Pdr}). 
If such a comparison can have been established, this would have explained:
\begin{itemize}

\item[(1)] why a particular MLWA is singled out from all possible MLWA approximations in the $x\ll 1$ range

\item[(2)] its success for $x\gtrsim 1$ (a kind of resummation)
(Note in passing that any fraction leads to infinite Taylor series in $x$.)

\end{itemize}

On introducing a shorthand $\Dt(x):= F + D(x) -i R(x)$ in eq 6,
together with the respective shorthands $d_\ell$ and $r_\ell$ for $x$-independent factors in $D(x)$ and $R(x)$,
one can verify that the only nonzero values and derivatives of $\Dt$, $D(x)$ and $R(x)$ in the limit $x\to 0$ are
\bea
& D^{(2)}(0)=2d_\ell,\qquad R^{(2\ell+1)}(0)=(2\ell+1)!\, r_\ell, &
\nn\\
&\Dt(0)=F, \qquad \Dt^{(2)}(0)=D^{(2)}(0)=2d_\ell, &
\nn\\
&\Dt^{(2\ell+1)}(0)=-iR^{(2\ell+1)}(0)=-i (2\ell+1)!\, r_\ell.&
\lb{ddrd}
\eea
In virtue of the general {\em Leibniz rule}, one finds for the $n$th derivative of $T_{E\ell}(x)$ of eq 6
\be
T^{(n)}_{E\ell}(0)=i \sum_{k=0}^n {n \choose k} R^{(k)}(0) [\Dt^{-1}(0)]^{(n-k)}, \nn
\eg
where ${n \choose k}$ is the familiar binomial coefficient. 
In virtue of \rfs{ddrd}
\bea
T^{(n)}_{E\ell}(0) &\equiv& 0, \qquad n=0,1,\ldots,2\ell,
\nn\\
T^{(2\ell+1)}_{E\ell}(0) &=& i R^{(2\ell+1)}(0) \Dt^{-1}(0) = \fr{i(2\ell+1)! r_\ell}{F},
\nn
\eea
whereas for $n>2\ell+1$
\bg
T^{(n)}_{E\ell}(0) = i {n \choose 2\ell+1 } R^{(2\ell+1)}(0) [\Dt^{-1}(0)]^{(n-1-2\ell)} = i {n \choose 2\ell+1 } (2\ell+1)!\, r_\ell \, [\Dt^{-1}(0)]^{(n-1-2\ell)}.
\lb{ng2lp1}
\eg
In order to determine $[\Dt^{-1}(0)]^{(k)}$, one can make use of the 
Fa\`a di Bruno's formula~\cite{Bruno1855,*Bruno1857}
\bg
\fr{d^n}{dx^n}\, f(g(x))=\sum \frac{n!}{m_1!\,m_2!\,\cdots \,m_n!}
\cdot f^{(m_{1}+\cdots +m_n)}(g(x))
\times
\prod_{j=1}^n \left(\frac{ g^{(j)}(x)}{j!}\right)^{m_j},
\lb{BBf}
\eg
where the sum is over all $n$-tuples of nonnegative integers $(m_1,\ldots, m_n)$ satisfying the constraint
$1\cdot m_1+2\cdot m_2+3\cdot m_3+\cdots +n\cdot m_n=\sum_{j=1}^n j m_j=n$ (obviously some of $m_j$'s will be zero).

The point of crucial importance is that, in virtue of \rfs{ddrd}, the rhs of \rfs{BBf}
is, in the case of $g=\Dt$, {\em nonzero} only 
for $j=2$ and $m_2=n/2$, i.e. requiring $n$ to be an {\em even} number and all 
other $m_j\equiv 0$, $j\ne 2$, $n<2\ell+1$. Therefore for $n<2\ell+1$ one finds
\bea
[\Dt^{-1}(x)]^{(n)} & = & \dfrac{n!}{(n/2)!} (-1)^{n/2} (n/2)!\, \Dt^{-(n/2)-1}(x) 
\left(\dfrac{\Dt^{(2)}(x)}{2!}\right)^{n/2}
\nn\\
&\to& (-1)^{n/2} n! F^{-(n/2)-1} d_\ell^{n/2} \qquad (x\to 0) \nn
\eea
On substituting the result back into \rfs{ng2lp1}, where $[\Dt^{-1}(0)]^{(n-1-2\ell)}$,
one finds for $2\ell+1<n\le 4\ell+1$ (i.e. $0\le n-1-2\ell<2\ell+1$) and $n$ {\em odd}
\bea
T^{(n)}_{E\ell}(0)
& = & i (-1)^{(n-1-2\ell)/2} {n \choose 2\ell+1 } (n-1-2\ell)!
(2\ell+1)!\, r_\ell
 F^{-((n-1-2\ell)/2)-1} d_\ell^{(n-1-2\ell)/2}
\nn\\
&=&
i (-1)^{(n-1-2\ell)/2} n! r_\ell
 F^{-((n-1-2\ell)/2)-1} d_\ell^{(n-1-2\ell)/2} , \nn
\eea
whereas $T^{(n)}_{E\ell}(0)\equiv 0$ for $n$ {\em even}.

Eventually, given that
\bg
[\Dt^{-1}(0)]^{(2\ell+1)}= i (2\ell+1)! F^{-2}\, r_\ell \qquad (x\to 0), \nn
\eg
one finds for {\em even} $n=2(2\ell+1)$
\bea
T^{(4\ell+2)}_{E\ell}(0) &=& i {4\ell+2 \choose 2\ell+1 } (2\ell+1)!\, r_\ell \, [\Dt^{-1}(0)]^{(2\ell+1)}
\nn\\
&=&
- {4\ell+2 \choose 2\ell+1 } F^{-2} [(2\ell+1)!\, r_\ell ]^2
\nn\\
&=&
- (4\ell+2)! F^{-2} r_\ell^2. \nn
\eea
Importantly, when comparing with the derivatives $T^{(n)}_{E\ell}(0)$ in the exact Mie theory,
one has to consider the variable $x_s=n_s x$ as dependent on $x$.

\bibliography{references}